\documentclass[aps,pre,preprint,showpacs,showkeys]{revtex4}
\usepackage{amsmath,amssymb}
\usepackage{graphicx,epsfig}

\begin{document}

\title{Two-time Green's functions and spectral density method in
nonextensive quantum statistical mechanics}
\author{A. Cavallo}
\affiliation{Institut Charles Sadron, Campus CNRS Cronenbourg, 23 rue du Loess,
BP 84047, 67034 Strasbourg Cedex 2, France.}
\author{F. Cosenza}
\email{cosfab@sa.infn.it}
\author{L. De Cesare}
\affiliation{Dipartimento di Fisica ''E.R. Caianiello'', Universit\`{a} degli Studi di
Salerno and CNISM, Unit\`{a} di Salerno, I-84081 Baronissi (SA), Italy.}
\homepage{www.physics.unisa.it}

\begin{abstract}
We extend the formalism of the thermodynamic two-time Green's
functions to nonextensive quantum statistical mechanics. Working
in the optimal Lagrangian multipliers representation, the $q$-spectral
properties and the methods for a direct calculation of the two-time $q$%
-Green's functions and the related $q$-spectral density ($q$ measures the
nonextensivity degree) for two generic operators are presented in strict
analogy with the extensive ($q=1$) counterpart. Some emphasis is devoted to
the nonextensive version of the less known spectral density method whose
effectiveness in exploring equilibrium and transport properties of a wide
variety of systems has been well established in conventional classical
and quantum many-body physics. To check how both the equations of motion and
the spectral density methods work to study the $q$-induced nonextensivity
effects in nontrivial many-body problems, we focus on the equilibrium
properties of a second-quantized model for a high-density Bose gas with
strong attraction between particles for which exact results exist in
extensive conditions. Remarkably, the contributions to several thermodynamic
quantities of the $q$-induced nonextensivity close to the extensive regime
are explicitly calculated in the low-temperature regime by overcoming the
calculation of the $q$ grand-partition function.
\end{abstract}

\maketitle



\section{Introduction\label{sec_1}}

The method of thermodynamic Green's functions (GFs) \cite%
{Abrikosov:65,Tyablikov:67,
Zubarev:74,kadanoff:62,mahan:90,majlis:00,kalashnikov:73,bogolyubov:63,cavallo:05, Abrikosov:91}
is a powerful tool in ordinary statistical mechanics for exploring the
equilibrium and transport properties of a large variety of many-body
systems. These functions are related to important physical quantities and
hence their calculation constitutes one of the basic problems of
extensive thermostatistics.

Remarkably, the extraordinary effectiveness of the GF technique in quantum
many-body physics \cite{Abrikosov:65,Tyablikov:67,
Zubarev:74,kadanoff:62,mahan:90,majlis:00,kalashnikov:73} has stimulated a lot
of research activity to extend this successful method also to study extensive
classical statistical mechanics \cite{bogolyubov:63,cavallo:05,Abrikosov:91}%
. In the context of the two-time GFs \cite{Tyablikov:67,Zubarev:74} the
foundations of the classical formalism were introduced by Bogoliubov and
Sadovnikov \cite{bogolyubov:63} four decades ago and further systematic
developments in this direction were performed only many years later \cite%
{Tyablikov:67,cavallo:05}. Additionally, the classical counterpart of the
quantum Matsubara GF framework \cite{mahan:90} was achieved in Ref. \cite%
{Abrikosov:91}. So also the two-time GF formalism in extensive classical
statistical mechanics can be now considered well established. Although
not currently used in the literature, it has been successfully employed to study the
thermodynamics and the transport properties of several classical many-body
systems \cite{Tyablikov:67,bogolyubov:63,cavallo:05,Abrikosov:91} also
involving phase transitions and critical phenomena \cite{cavallo:05}.

Various (also numerical) methods have been developed for the calculation of the
two-time GFs both in quantum and classical statistical physics \cite%
{Abrikosov:65,Tyablikov:67,
Zubarev:74,kadanoff:62,mahan:90,majlis:00,kalashnikov:73}. In addition to the most
known \textquotedblleft equations-of-motion method\textquotedblright\ (EMM),
the related \textquotedblleft spectral density method
\textquotedblright\ (SDM), originally formulated by Kalashnikov and Fradkin
\cite{kalashnikov:73} within the quantum statistical mechanics context,
appears to be a very promising nonperturbative approach to perform reliable
studies of the macroscopic properties of classical and quantum many-particle
systems \cite{kalashnikov:73,cavallo:05} avoiding an explicit calculation
of the partition function.

At the present time, the situation does not appear so well established in nonextensive statistical mechanics which has
attracted an increasing interest since the seminal proposal by Tsallis made almost
20 years ago \cite{tsallis:88,tsallis:98}.
This framework can be regarded as a
generalization of the Boltzmann-Gibbs statistical mechanics to properly
describe macroscopic memory effects (e.g., non-Markovian stochastic processes)
and, generally speaking, systems exhibiting nonergodic microscopic
dynamics \cite{2001,NEXT rev,web}.

Despite its elegant formalism, the Tsallis thermostatistics is affected by
some intrinsic difficulties in performing explicit analytical calculations
for realistic many-body systems. Nevertheless, by resorting to different
techniques and approximations as generalizations of the extensive ones, the
Tsallis theory has been successfully applied to a wide variety of systems
for which nonextensivity effects are not negligible \cite{2001,NEXT rev,web}
and must be taken into account for a proper comparison with experiments.
Known theoretical tools have been employed and adapted in the Tsallis
framework such as, for instance, linear response theory \cite{rajagopal:96},
perturbation and variational methods \cite{rajagopal:98}, path integral
\cite{Ex18,Ex19}, Monte Carlo \cite%
{tsallis:96,straub:99,andricioaei:96,lima:99,salazar:99} and molecular
dynamics \cite{straub:99} techniques and many others. Less attention,
however, has been devoted to the extension, to the nonextensive many-body
world, of the well-established thermodynamic GF technique \cite%
{Abrikosov:65,Tyablikov:67,Zubarev:74,kadanoff:62,
mahan:90,majlis:00,kalashnikov:73,bogolyubov:63,cavallo:05, Abrikosov:91} in
 ordinary quantum and classical statistical mechanics.

Some years ago the GF method in the Kadanoff-Baym framework \cite%
{kadanoff:62} was generalized to the Tsallis quantum statistical mechanics
adopting the second quantized representation for many-particle systems \cite%
{Lenzi:99,Abe:99}. In these works, the $q$ GFs for a nonextensive many-body
system were formally expressed in terms of parametric integrals over the
corresponding extensive $(q=1)$ quantities, $q$ denoting the so-called
Tsallis parameter which measures the nonextensivity degree. Of course, this
method may be really useful when the many-body problem for the extensive
counterpart has been solved but, unfortunately, this is possible only in a
limited number of situations. In any case, the crucial step to calculate
nontrivial contour integrals in the complex space constitutes an additional
formidable problem which would require, generally, further approximations.
Thus, many-body methods which allow one to perform direct calculations of the $q$%
properties, overcoming the \emph{a priori} knowledge of the related extensive
ones, are desirable. Motivated by the conviction that a direct $q$ GF
method may provide new and effective calculation techniques to deal with
nontrivial nonextensivity problems, we have recently extended \cite%
{cavallo:01,cavallo:02} the Bogoliubov-Sadovnikov two-time GF framework
\cite{bogolyubov:63,cavallo:05} and the related SDM \cite{cavallo:05} to
nonextensive classical statistical mechanics working conveniently within the
so-called optimal Lagrangian multipliers (OLM) representation suggested in
Ref. \cite{martinez:00}. This choice avoids some intrinsic difficulties involved in other
ones and allows simplified analytical and numerical calculations. In any
case, our suggestion is quite general and can be extended to different
contexts preserving the physical content, consistently with the equivalence
of the current four versions of the Tsallis statistics \cite{Ferri:05}.

The aim of the present article is to extend the two-time GF formalism to
the quantum nonextensive statistical mechanics within the OLM representation.
Particular emphasis will be devoted to the spectral density (SD) and its spectral decomposition due to their relevance for practical calculations.

The $q$ EMM and the $q$ SDM are here presented as powerful tools for a
direct calculation of the two-time $q$ GFs and $q$ SD, respectively.
Besides, in parallel with the extensive counterpart \cite{cavallo:05}, we outline
the key ideas to explore the dispersion relation and the damping of
elementary and collective excitations in nonextensive many-body systems.
Finally, with the aim to show how the two-time GF method works when the
Tsallis $q$ distribution is involved, we present analytical calculations for
a nontrivial high-density Bose gas with strong attraction between the
particles for which exact results exist in the extensive case \cite%
{Babichenko:73,Campana_NC:79}. The effects of the $q$-induced nonextensivity
will be consistently explored in the low-temperature regime using both the $%
q $ EMM and $q$ SDM.

The article is organized as follows. In Sec. \ref{sec_2}, we present a
summary of the basic ingredients of the nonextensive quantum statistical
mechanics which will be useful for the next developments. In particular, we
focus on the quantum OLM representation which appears more convenient for
our purposes. Section \ref{sec_3} is devoted to the formulation of the
two-time GF method in the Tsallis quantum statistics, with emphasis on the
spectral properties. In Sec. \ref{sec_4}, we present a nonextensive
version of the EMM and SDM. Here, the $q$ SDM is properly implemented to
offer the possibility to study systematically the dispersion relation and
the damping of excitations within a unified formalism. Section \ref{sec_5}
deals with the application of these methods to a model which describes a
high density Bose gas with strong attraction between the particles.
Concluding remarks are drawn in Sec. \ref{sec_6}. Two appendixes close the
article. In Appendix A the recently proposed \cite{cavallo:01,cavallo:02}
two-time GF method in Tsallis classical statistical mechanics is shortly
reviewed with the aim to point out the substantial differences between the
nonextensive quantum and classical frameworks. Appendix B summarizes
some mathematical details.


\section{Basic ingredients of nonextensive quantum statistical mechanics
\label{sec_2}}

The Tsallis nonextensive thermostatistics constitutes today a new paradigm in
the field of statistical mechanics. The key problem of the Tsallis framework
is to find the appropriate von Neumann density operator $\rho $ which
maximizes the generalized $q$ entropy (in units $k_{B}=\hbar =1$)
\begin{equation}
S_{q}=\frac{1-T_{r}\left( \rho ^{q}\right) }{q-1},  \label{eq:GqE}
\end{equation}%
subject to appropriate constraints related to the evaluation of the mean or
expectation value of the observables within the nonextensive scenario. Here $%
T_{r}(\ldots )$ stands for the usual trace operator.

In the literature, four possible choices have been considered for the
evaluation of $q$ expectation values $\langle ...\rangle _{q}$:

\begin{itemize}
\item[(i)] The original Tsallis proposal \cite{tsallis:88}
\begin{equation}
\langle A\rangle _{q}=T_{r}(\rho A),  \label{eq:oTp}
\end{equation}%
where the Hermitian operator $A$ corresponds to a generic observable $%
\mathcal{A}$. For a system with Hamiltonian $H$, this implies the canonical
representation
\begin{equation}
\rho =\left[ \left( 1-q\right) \left( \alpha +\beta H\right) /q\right] ^{%
\frac{1}{q-1}},
\end{equation}%
in terms of the two Lagrange multipliers $\alpha $ and $\beta $.
Unfortunately, this choice involves some troubles related to the Lagrange multiplier $%
\alpha $.

\item[(ii)] The Curado-Tsallis (CT) choice \cite{Curado:91}%
\begin{equation}
\langle A\rangle _{q}=T_{r}(\rho ^{q}A),  \label{eq:CT}
\end{equation}%
which yields the canonical result
\begin{equation}
\rho =Z_{q}^{-1}\left[ 1-\left( 1-q\right) \beta H\right] ^{\frac{1}{1-q}},
\label{eq:CR1}
\end{equation}%
with
\begin{equation}
Z_{q}=T_{r}\left[ 1-\left( 1-q\right) \beta H\right] ^{\frac{1}{1-q}}.
\label{eq:CR2}
\end{equation}%
This avoids the explicit presence of the multiplier $\alpha $ but has the
disadvantage to exhibit unnormalized mean values ($\langle1\rangle _{q}\neq
1 $).

\item[(iii)] The Tsallis-Mendes-Plastino option \cite{tsallis:98}
\begin{equation}
\langle A\rangle _{q}=\frac{T_{r}\left( \rho ^{q}A\right) }{T_{r}(\rho ^{q})}%
,  \label{eq:TMP}
\end{equation}%
implying the canonical solution
\begin{equation}
\rho =Z_{q}^{-1}\left[ 1-\frac{\left( 1-q\right) \beta }{T_{r}\left( \rho
^{q}\right) }\left( H-U_{q}\right) \right] ^{\frac{1}{1-q}},  \label{eq:TMP1}
\end{equation}%
with
\begin{equation}
Z_{q}=T_{r}\left[ 1-\frac{\left( 1-q\right) \beta }{T_{r}\left( \rho
^{q}\right) }\left( H-U_{q}\right) \right] ^{\frac{1}{1-q}},  \label{eq:TMP2}
\end{equation}%
where $U_{q}=\langle H\rangle _{q}$ is the $q$ internal energy. Here we have
$\langle 1\rangle _{q}=1$ normalized $q$ mean values, but troubles occur
again in obtaining $\rho $ due to the presence of $T_{r}\left( \rho
^{q}\right) \equiv \left( Z_{q}\right) ^{1-q}$ in Eqs.(\ref{eq:TMP1}) and (\ref%
{eq:TMP2}) (self-referential problem).

\item[(iv)] The OLM improvement \cite{martinez:00} which preserves the $q$%
 entropy (\ref{eq:GqE}), but replaces the Tsallis-Mendes-Plastino-like
constraints by \textquotedblleft centered\textquotedblright\ mean values.
Essentially, the general OLM procedure consists in maximizing the Tsallis
generalized entropy (\ref{eq:GqE}) subject to the constraints
\begin{equation}
T_{r}\rho =1  \label{eq:constraint1}
\end{equation}%
and
\begin{equation}
T_{r}\left[ \rho ^{q}\left( A_{j}-\langle A_{j}\rangle _{q}\right) \right]
=0 \;\; (j=1\ldots ,M),  \label{eq:constraint2}
\end{equation}%
where the generalized mean values $\langle A_{j}\rangle _{q}=T_{r}\left(
\rho ^{q}A_{j}\right) /T_{r}(\rho ^{q})$ for $M$ relevant $A_{j}$ ($j=1,...,M$%
) observables are assumed to be known a priori and hence are regarded as
constraints in the variational approach to the nonextensive thermostatistics.

In the canonical representation this yields
\begin{equation}
\rho =Z_{q}^{-1}\left[ 1-\left( 1-q\right) \beta \left( H-U_{q}\right) %
\right] ^{\frac{1}{1-q}},  \label{eq:CRrho}
\end{equation}%
where the generalized partition function $Z_{q}$ is now given by
\begin{equation}
Z_{q}=T_{r}\left[ 1-\left( 1-q\right) \beta \left( H-U_{q}\right) \right] ^{%
\frac{1}{1-q}}.  \label{eq:GPF}
\end{equation}%
As we see, the OLM framework avoids all the inconveniences occurring in the
previous choices. Besides, in Eqs. (\ref{eq:CRrho}) and (\ref{eq:GPF}), the
Lagrange multiplier $\beta $ does not depend on the partition function and
it is identified as the inverse physical temperature \cite%
{martinez:00,Ferri:05} $\left( \beta =1/T\right) $, consistently with the
zeroth law of thermodynamics.

The extension of the canonical OLM prescription to the grand-canonical
ensemble can be simply obtained from Eqs. (\ref{eq:CRrho}) and (\ref{eq:GPF})
by replacing $H$ and $U_{q}$ with $\mathcal{H}=H-\mu N$ and $\mathcal{U}%
_{q}=\langle \mathcal{H}\rangle _{q}=U_{q}-\mu \langle N\rangle _{q}$, where
$\mu $ is the chemical potential and $N$ denotes the operator describing the
total number of particles in a system.
Other ensemble representations can be obtained similarly.
\end{itemize}

It must be emphasized that the four versions (i)-(iv) of the Tsallis
thermostatistics are equivalent in the sense that the probability
distribution for each of them can be easily derived from any other of them
by using appropriate transformation rules \cite{Ferri:05}. In any case, for
practical calculations one must select the most convenient version case by
case.

As already mentioned, in the present work we find it convenient to adopt the
OLM version to formulate the two-time GF method in nonextensive
statistical mechanics, but the framework can be easily extended to different
versions by using the prescriptions given in Ref. \cite{Ferri:05}.

To close this short review of the basic elements of the Tsallis
thermostatistics, we mention two relevant features of
the nonextensive scenario.

The first one, tacitly assumed before, is that the basic OLM canonical
operator $\widehat{f}_{q}=1-(1-q)\beta (H-U_{q})$
must be positive definite. This means that, for any specific nonextensive problem,
one must take into account only microstates $\left\vert n\right\rangle $,
with $H\left\vert n\right\rangle =E_{n}\left\vert n\right\rangle $,
satisfying the cutoff condition \cite{tsallis:98}
\begin{subequations}
\label{eq:cond_13}
\begin{equation}
E_{n}<\frac{1}{(1-q)\beta }+U_{q}\text{ \ , if }q<1,
\end{equation}
\begin{equation}
E_{n}>\frac{1}{(1-q)\beta }+U_{q}\text{ \ , if }q>1,%
\end{equation}%
\end{subequations}
where the quantity $1/(1-q)\beta +U_{q}$ could have, in principle, positive
or negative sign.

The second crucial question concerns the physical origin of the
nonextensivity parameter $q$. Unfortunately, little has been definitevely
proposed in terms of first principles to explain the appearance of the
Tsallis thermostatistics in many real situations and to understand the
physical meaning of the elusive parameter $q$. We cite below
some general ideas which, in our opinion, may have a seminal role in future
research activity on the subject.

After several attempts \cite{web}, a well
definited scenario emerged almost seven years ago,
now called the "finite-heat-bath picture".
First Almeida \cite{Almeida:01} derived
the Tsallis power-law probability distribution from first principles assuming
exactly constant the heat capacity $C_{B}$ of the thermal bath in contact
with the system of interest. Here, the
physical meaning of $q$ is simply expressed in terms of $C_{B}$.
Specifically, when it is finite, $q\neq 1$ and one has the Tsallis
distribution, while for an infinite heat capacity of the bath, $q=1$ and
one recovers the conventional exponential distribution. This picture has been
further elaborated \cite{Andrade:etc,Tasuaki:03}, but a substantial progress was
performed in Ref. \cite{Tasuaki:03} where, via a model-free derivation of the Tsallis
statistics (i.e., without resort to the microscopic details of a system and
of its surrounding), it is shown that the finiteness of the heat capacity of the
environment is not necessarily due to only the finiteness of its
degrees of freedom, as reductively assumed in previous treatments.

A more general point of view was reported in Ref. \cite{Wilk:00}
where the parameter $q$ was interpreted as a measure of fluctuations of the
parameters (as the temperature) in the exponential distribution and
expressed in terms of the variance of their inverse.
The superstatistics \cite{Beck:06}, subsequently developed as a
generalization of this idea, consists in a superposition of different statistics relevant
for driven nonequilibrium systems with complex dynamics in stationary states
with large fluctuations of intensive parameters (e.g., the temperature,
chemical potential, energy dissipation, etc.) on long time scales. It
explains the emergence of the power-law statistics for real systems as a
result of fluctuations in their environments, supporting the Tsallis
framework as a special case. In this "fluctuational picture," the
nonextensivity parameter $q$ is found to be directly related to proper
stochastic processes or constraints imposed to the systems and,
in the absence of fluctuations, the extensive case is consistently
reproduced. Remarkably, the superstatistics seems to offer a general formalism
for treating nonequilibrium stationary states of complex systems which exhibit
dynamics that can be decomposed into several dynamics on different time scales.
Some effort has been performed recently \cite{ref38bis} to legitimate this theory as a true
statistical mechanics framework, but several aspects remain to be further elaborated
and clarified.

Other, rather fragmentary, proposals to justify the Tsallis statistics
exist, but we mention here only some of them that, in our opinion, may be relevant
in nonextensive many-body physics.

There is a broad consensus of opinion that the nonextensivity may occur also in cases where
long-range interactions play a relevant role, the gravitational and Coulomb
forces being important examples with laboratory and astrophysics
implications \cite{web}. However, no definitive statistical
foundation for long-range interacting systems, such as the self-gravitating
ones, has still been well established and, in particular, the true physical
nature of $q$ in terms of the range of the forces and the space dimensions
has not been well understood yet \cite{Du:06}.

Also confining traps for interacting bosonic or fermionic systems may induce
nonextensivity effects \cite{Tonar:02} due to the
interplay between the particle interactions and the trapping potential.
Moreover, interesting findings about the physical origin and the
measurement of the parameter $q$ have been obtained recently for complex
magnetic systems such as the manganite $La_{0.7}Sr_{0.3}MnO_{3}/MgO$ and the
amorphous alloy $Cu_{90}Co_{10}$ \cite{Oliveira:05}.
Such compounds seem to embody three basic ingredients which may
induce nonextensivity: long-range interactions, clusters with fractal shapes,
and intrinsic inhomogeneity. With these features in mind it has been shown \cite{Oliveira:05},
via theory (at a mean
-field approximation level) and scanning tunneling
spectroscopy measurements, that the magnetic properties of these materials
can be properly described using the Tsallis thermostatistics, $q$ measuring
the competition of the intrinsic inhomogeneity and dynamics.

Finally,
for our demonstrative application to a high-density Bose
gas performed in a next section, we
have in mind some of the nonextensivity mechanisms outlined before. For
instance, we think about a system of bosons in a fluctuating
environment, in confining traps, or confined to a self-gravitational field as
in non-relativistic boson stars \cite{Jetzer:92}.

We conclude this section noting that it is frequently convenient to use the
so-called $q$ exponential function
\begin{equation}
e_{q}^{x}\equiv \left[ 1+\left( 1-q\right) x\right] ^{\frac{1}{1-q}},
\label{eq:qEF}
\end{equation}%
which simplifies sensibly the formalism of the nonextensive statistical
mechanics. One can verify that this function (a) yields $e_{1}^{x}=e^{x}$
for $q\rightarrow 1$, (b) for $q>1$ vanishes as a power law when $%
x\rightarrow -\infty$, and (c) diverges at $x=-1/\left( 1-q\right) $; and 3) for
$q<1$ has a cutoff at $x=-1/\left( 1-q\right) $, below which it becomes
identically zero.

Using the function (\ref{eq:qEF}), the basic expressions (\ref{eq:CRrho})
and (\ref{eq:GPF}) for the OLM canonical formulation can be rewritten as
\begin{equation}
\rho =Z_{q}^{-1}e_{q}^{-\beta \left( H-U_{q}\right) },  \label{eq:exOLM1}
\end{equation}%
and
\begin{equation}
Z_{q}=T_{r}e_{q}^{-\beta \left( H-U_{q}\right) }.  \label{eq:exOLM2}
\end{equation}%
It is then immediate to see that, for $q\rightarrow 1$, Eqs. (\ref{eq:exOLM1}%
) and (\ref{eq:exOLM2}) reproduce the conventional Boltzmann-Gibbs framework.

\section{Two-time Green's functions and spectral density in quantum
nonextensive statistical mechanics\label{sec_3}}

\subsection{Definitions and spectral properties\label{sec_3.1}}

In strict analogy with the extensive case \cite%
{Tyablikov:67,Zubarev:74,majlis:00}, we define the two-time retarded $(\nu
=r)$ and advanced $\left( \nu =a\right) $ $q$ GFs in quantum nonextensive
thermostatistics for two operators $A$ and $B$ as \cite{nota:app.A}
\begin{eqnarray}
G_{qAB}^{(\nu )}(t,t^{\prime }) &=&-i\theta _{\nu }\left( t-t^{\prime
}\right) \langle \left[ A(t),B(t^{\prime })\right] _{\eta }\rangle _{q}
\notag \\
&\equiv &\langle \langle A(t);B(t^{\prime })\rangle \rangle _{q}^{(\nu )},\
\ \left( \nu =r,a\right) .  \label{eq:QNESM}
\end{eqnarray}%
Here, $\theta _{r}\left( t-t^{\prime }\right) =\theta \left( t-t^{\prime
}\right) $, $\theta _{a}\left( t-t^{\prime }\right) =-\theta \left(
t^{\prime }-t\right) $, and $\theta (x)$ is the step function. In Eq. (\ref%
{eq:QNESM}), $\left[ ... ,... \right] _{\eta }$ denotes a commutator $\left(
\eta =-1\right) $ or an anticommutator $\left( \eta =+1\right) $ and $%
X(t)=e^{iHt}Xe^{-iHt}$ is the Heisenberg representation of the operator $X$,
satisfying the Heisenberg equation of motion (EM) \cite{nota:commutatori}
\begin{equation}
\frac{dX(t)}{dt}=i\left[ H,X(t)\right] _{-}.  \label{eq:HE}
\end{equation}%
The definition (\ref{eq:QNESM}) reproduces the conventional extensive
formalism for $q=1$ and allows us to develop the $q$ GF framework
equivalently with commutators or anticommutators. However, in practical
calculations it will be convenient to use, in Eq. (\ref{eq:QNESM}), $\eta
=-1 $ or $\eta =+1$ for bosonic or fermionic operators, respectively.

Physically relevant quantities, which enter the definition of $q$ GFs, are
the two time $q$ correlation functions (CFs) $F_{qAB}(t,t^{\prime
})=\langle A(t)B(t^{\prime })\rangle _{q}$ and $F_{qBA}(t^{\prime},t)=%
\langle B(t^{\prime })A(t)\rangle _{q}$ for the corresponding operators.
Working within the equilibrium statistics, one can easily prove that the
two-time $q$ CFs and $q$ GFs depend on times $t$ and $t^{\prime }$ only
through the difference $\tau =t-t^{\prime }$. So one can write
\begin{equation}
G_{qAB}^{(\nu )}(t-t^{\prime })=\langle \langle A(t-t^{\prime });B\rangle
\rangle _{q}^{(\nu )}=\langle \langle A;B(t^{\prime }-t)\rangle \rangle
_{q}^{(\nu )}.  \label{eq:QNESM2}
\end{equation}%
This feature allows us to introduce the Fourier transforms
\begin{equation}
G_{qAB}^{(\nu )}(\tau )=\int_{-\infty }^{+\infty }\frac{d\omega }{2\pi }%
G_{qAB}^{(\nu )}(\omega )e^{-i\omega \tau },  \label{eq:FT1}
\end{equation}%
\begin{equation}
F_{qXY}(\tau )=\int_{-\infty }^{+\infty }\frac{d\omega }{2\pi }%
F_{qXY}(\omega )e^{-i\omega \tau },  \label{eq:FT2}
\end{equation}%
where $G_{qAB}^{(\nu )}(\omega )=\langle \langle A(\tau );B\rangle \rangle
_{q,\omega }^{(\nu )}$ will be named the $\nu $-type $q$ GF of $A$ and $B$
in the $\omega $-representation and $F_{qXY}(\omega )\equiv \langle X(\tau
)Y\rangle _{q,\omega }$ will be called the $q$-spectral intensity of the
time-dependent $q$ CF $F_{qXY}(\tau )$, with $\Im (\omega
)=\int_{-\infty }^{+\infty }d\tau \Im (\tau )e^{i\omega \tau }$ $\left( \Im
=G_{qAB}^{(\nu )},F_{qAB},\ldots \right) $.

We now define the time-dependent $q$ SD for the operators $A$ and $B$ \cite%
{Tyablikov:67,Zubarev:74,kalashnikov:73} as
\begin{equation}
\Lambda _{qAB}(\tau )=\langle \left[ A(\tau ),B\right] _{\eta }\rangle _{q}.
\label{eq:TdqSD}
\end{equation}%
For its Fourier transform (the $q$ SD in the $\omega $-representation)
\begin{equation}
\Lambda _{qAB}(\omega )=\langle \left[ A(\tau ),B\right] _{\eta }\rangle
_{q,\omega },  \label{eq:TdqSD_FT}
\end{equation}%
one immediately finds the exact result
\begin{equation}
\int_{-\infty }^{+\infty }\frac{d\omega }{2\pi }\Lambda _{qAB}(\omega
)=\langle \left[ A,B\right] _{\eta }\rangle _{q},  \label{eq:TdqSD_ER}
\end{equation}%
which constitutes an important \textquotedblleft sum rule\textquotedblright\
for the $q$ SD $\Lambda _{qAB}(\omega )$ to be used for physical consistency
of practical calculations and approximations.

Then, using the integral representation for the $\nu $-step function
\begin{equation}
\theta _{\nu }(\tau )=i\int_{-\infty }^{+\infty }\frac{dx}{2\pi }\frac{%
e^{-ix\tau }}{x+(-1)^{\nu }i\varepsilon },\ \ \varepsilon \rightarrow 0^{+},
\label{eq:IR_ST}
\end{equation}%
where the symbol $(-1)^{\nu }$ stands for $+1$ if $\nu =r$ and $-1$ if $\nu
=a$, we obtain the $q$-spectral representation
\begin{equation}
G_{qAB}^{(\nu )}(\omega )=\int_{-\infty }^{+\infty }\frac{d\omega ^{\prime }%
}{2\pi }\frac{\Lambda _{qAB}(\omega ^{\prime })}{\omega -\omega ^{\prime
}+(-1)^{\nu }i\varepsilon },\ \ \varepsilon \rightarrow 0^{+},
\label{eq:qSR}
\end{equation}%
for the Fourier transforms of the two-time $q$ GFs (\ref{eq:QNESM2}).

Combining Eq. (\ref{eq:qSR}) and the sum rule (\ref{eq:TdqSD_ER}), one can
get another general result which, as in the extensive case, may play a
relevant role in practical calculations involving the $q$ GFs. Indeed, as $%
|\omega |\rightarrow \infty $ we have
\begin{multline}
G_{qAB}^{(\nu )}(\omega ) = \omega ^{-1}\int_{-\infty }^{+\infty }\frac{%
d\omega ^{\prime }}{2\pi }\frac{\Lambda _{qAB}(\omega ^{\prime })}{1-\frac{%
\omega ^{\prime }-(-1)^{\nu }i\varepsilon }{\omega }} \approx \\
\approx \frac{\langle %
\left[ A,B\right] _{\eta }\rangle _{q}}{\omega } 
+\frac{1}{\omega ^{2}}\int_{-\infty }^{+\infty }\frac{d\omega ^{\prime }}{%
2\pi }\Lambda _{qAB}(\omega ^{\prime })\left[ \omega ^{\prime }-(-1)^{\nu
}i\varepsilon \right] +O\left( \frac{1}{\omega ^{3}}\right) ,
\label{eq:qSR_gr}
\end{multline}%
and hence
\begin{equation}
G_{qAB}^{(\nu )}(\omega )\sim \left\{
\begin{array}{ll}
\omega ^{-1},\ \ \mathrm{if}\ \ \langle \left[ A,B\right] _{\eta }\rangle
_{q}\neq 0 &  \\
\omega ^{-\alpha },\ \ (\alpha \geq 2),\ \ \mathrm{if}\ \ \langle \left[ A,B%
\right] _{\eta }\rangle _{q}=0, &
\end{array}%
\right.  \label{eq:Gq_asymp}
\end{equation}%
which provides a relevant boundary condition for the $q$ GFs.

As in the quantum extensive case \cite{Tyablikov:67,Zubarev:74} one can
easily show that the $q$ GFs $G_{qAB}^{(\nu )}(\omega )$, analytically
continued in the complex $\omega $ plane, are analytical functions in the
upper (for $\nu =r$) and lower (for $\nu =a$) half planes. Then, these
functions can be combined to construct the $q$ GF of complex $\omega $:
\begin{equation}
G_{qAB}(\omega ) =\int_{-\infty }^{+\infty }\frac{d\omega ^{\prime }}{2\pi }%
\frac{\Lambda _{qAB}(\omega )}{\omega -\omega ^{\prime }} 
=\left\{
\begin{array}{ll}
G_{qAB}^{(r)}(\omega ),\ \ Im\omega >0 &  \\
G_{qAB}^{(a)}(\omega ),\ \ Im\omega <0, &
\end{array}%
\right.  \label{eq:qSR_gr2}
\end{equation}
which is analytical in the whole complex $\omega $ plane with a cut along
the real axis where singularities may occur.

It is worth noting that, in terms of the SD, no formal differences exist for
the spectral representations of the GFs in the extensive and nonextensive
contexts. Hence, most of the developments in the extensive two-time GF
framework remain formally valid in the nonextensive one. For instance, using
the $\delta $-function representation
\begin{equation}
\delta (x)=\lim_{\varepsilon \rightarrow 0^{+}}\frac{1}{2\pi i}\left\{ \frac{%
1}{x-i\varepsilon }-\frac{1}{x+i\varepsilon }\right\} ,  \label{eq:deltaf}
\end{equation}%
we have the relation
\begin{equation}
\Lambda _{qAB}(\omega ) = i\left[ G_{qAB}\left( \omega +i\varepsilon
\right) -G_{qAB}\left( \omega -i\varepsilon \right) \right]
= i\left[ G_{qAB}^{(r)}\left( \omega \right) -G_{qAB}^{(a)}\left( \omega
\right) \right] ,  \label{eq:qSD_omegarep}
\end{equation}%
which expresses the $q$ SD in terms of the related two-time $q$ GFs in the $%
\omega$ representation. This result, which is expected to play an important
role in the applications of the $q$ GF method (as happens in the
extensive case), suggests also that the cut for $G_{qAB}(\omega )$ along the
real axis is determined by Eq. (\ref{eq:qSD_omegarep}) and its singularities
are given by real $\omega $ values satisfying the condition $\Lambda
_{qAB}(\omega )\neq 0$. It is worth noting that, in general,
$G_{qAB}(\omega )$ has to be regarded as a many-valued function of the
complex variable $\omega $. Hence, its singularities lie on the real axis on
the first Riemann sheet. On the other sheets, the singularities may shift to
the complex plane, leading to the appearance of complex poles.

Another important result can be easily obtained assuming $\Lambda
_{qAB}(\omega )$ real and using the relation
\begin{equation}
\lim_{\varepsilon \rightarrow 0^{+}}\int_{-\infty }^{+\infty }d\omega
^{\prime }\frac{f(\omega ^{\prime })}{\omega ^{\prime }-\omega +(-1)^{\nu
}i\varepsilon }
=\mathcal{P}\int_{-\infty }^{+\infty }d\omega ^{\prime }\frac{f(\omega
^{\prime })}{\omega ^{\prime }-\omega }-(-1)^{\nu }i\pi f(\omega ),
\label{eq:relation}
\end{equation}%
where $\mathcal{P}$ denotes the main part of the integral. From Eq.(\ref%
{eq:qSR}), we obtain indeed
\begin{equation}
G_{qAB}^{(\nu )}(\omega )=-\mathcal{P}\int_{-\infty }^{+\infty }\frac{%
d\omega ^{\prime }}{2\pi }\frac{\Lambda _{qAB}(\omega ^{\prime })}{\omega
^{\prime }-\omega }-\frac{(-1)^{\nu }}{2}i\Lambda _{qAB}(\omega ).
\label{eq:G_p}
\end{equation}%
Hence we get
\begin{equation}
\texttt{Re}G_{qAB}^{(\nu )}(\omega )=(-1)^{\nu }\mathcal{P}\int_{-\infty }^{+\infty }%
\frac{d\omega ^{\prime }}{\pi }\frac{\texttt{Im}G_{qAB}^{(\nu )}(\omega ^{\prime })}{%
\omega ^{\prime }-\omega },  \label{eq:realG_p}
\end{equation}%
with
\begin{equation}
\Lambda _{qAB}(\omega )=-2(-1)^{\nu }\texttt{Im}G_{qAB}^{(\nu )}(\omega ),
\label{eq:lambda1}
\end{equation}%
and, in particular,
\begin{equation}
\Lambda _{qAB}(\omega )=-2\texttt{Im}G_{qAB}^{(r)}(\omega ).  \label{eq:lambda2}
\end{equation}%
In analogy with the extensive counterparts \cite{Tyablikov:67,Zubarev:74},
the relations (\ref{eq:realG_p}) between the real and imaginary parts of
$G_{qAB}^{(r)}(\omega )$ and $G_{qAB}^{(a)}(\omega )$ will be called $q$%
 dispersion relations or nonextensive Kramer-Kronig relations.

In the next subsection we will derive spectral decompositions for $\Lambda
_{qAB}(\omega )$, $G_{qAB}(\omega )$, $F_{qAB}(\tau )$ and $F_{qAB}(\omega )$
which allow us to obtain information about the nature of the GF singularities and hence
about the excitations in nonextensive quantum many-body systems.

\subsection{$q$-spectral decompositions\label{sec_3.2}}

Let $\left\{ |n\rangle \right\} $ and $\left\{ E_{n}\right\}$ be the selected
eigenvectors and eigenvalues of the Hamiltonian $H$ of a many-body
system and assume that $\left\{ |n\rangle \right\} $ is a complete
orthonormal set of states. In this representation, the $q$ SD $\Lambda
_{qAB}(\omega )$, as given by the Fourier transform (\ref{eq:TdqSD_FT}), can
be written as
\begin{multline}
\Lambda _{qAB}(\omega )=\frac{2\pi }{\widetilde{Z}_{q}}\sum_{n,m}\left[
1-\left( 1-q\right) \beta \left( E_{n}-U_{q}\right) \right] ^{\frac{q}{1-q}%
} \times \\ \times
\left\{ 1+\eta \left[ \frac{1-\left( 1-q\right) \beta \left(
E_{m}-U_{q}\right) }{1-\left( 1-q\right) \beta \left( E_{n}-U_{q}\right) }%
\right] ^{\frac{q}{1-q}}\right\} 
 A_{nm}B_{mn}\delta \left( \omega -\omega _{mn}\right) ,
\label{eq:L_SD}
\end{multline}%
where use is made of the OLM canonical framework for calculation of the $q$%
 averages. In Eq. (\ref{eq:L_SD}),
\begin{equation}
\widetilde{Z}_{q}=T_{r}\rho ^{q}=\sum_{n}\left[ 1-\left( 1-q\right) \beta
\left( E_{n}-U_{q}\right) \right] ^{\frac{q}{1-q}},
\end{equation}%
$X_{nm}=\langle n|X|m\rangle $, and $\omega _{mn}=E_{m}-E_{n}$. Besides, the
spectral representation (\ref{eq:qSR_gr2}) for $G_{qAB}(\omega )$ yields
\begin{multline}
G_{qAB}(\omega )=\frac{2\pi }{\widetilde{Z}_{q}}\sum_{n,m}\left[ 1-\left(
1-q\right) \beta \left( E_{n}-U_{q}\right) \right] ^{\frac{q}{1-q}}\times \\
\times \left\{ 1+\eta \left[ \frac{1-\left( 1-q\right) \beta \left(
E_{m}-U_{q}\right) }{1-\left( 1-q\right) \beta \left( E_{n}-U_{q}\right) }%
\right] ^{\frac{q}{1-q}}\right\} \frac{A_{nm}B_{nm}}{\omega -\omega _{mn}}.
\label{eq:G_SD}
\end{multline}%
Analogously, for the two-time $q$-CF $F_{qAB}(\tau )$ and its Fourier
transform $F_{qAB}(\omega )$, one easily finds
\begin{equation}
F_{qAB}(\tau )=\frac{2\pi }{\widetilde{Z}_{q}}\sum_{n,m}\left[ 1-\left(
1-q\right) \beta \left( E_{n}-U_{q}\right) \right] ^{\frac{q}{1-q}} 
 A_{nm}B_{mn}e^{-i\omega _{mn}\tau }  \label{eq:F_SD1}
\end{equation}%
and
\begin{equation}
F_{qAB}(\omega ) = \frac{2\pi }{\widetilde{Z}_{q}}\sum_{n,m}\left[ 1-\left(
1-q\right) \beta \left( E_{n}-U_{q}\right) \right] ^{\frac{q}{1-q}}
A_{nm}B_{mn}\delta \left( \omega -\omega _{mn}\right) ,
\label{eq:F_SD2}
\end{equation}%
with similar expressions for $F_{qBA}(\tau )$ and $F_{qBA}(\omega )$.

The comparison of the previous relations with the corresponding extensive ones \cite%
{Tyablikov:67,Zubarev:74} indicates that the Tsallis statistics does not
modify the meaning of the GF singularities, but changes substantially the
structure of the spectral weights with the introduction of a mixing of the
energy levels which is absent in the extensive framework. Equations (\ref%
{eq:G_SD}) and (\ref{eq:F_SD1}) suggest indeed that the real poles of $%
G_{qAB}(\omega )$ (i.e., the frequencies $\omega _{mn}$) which are related to
the eigenvalues of the Hamiltonian, represent the frequency (energy)
spectrum of undamped excitations [oscillations in time of $F_{qAB}(\tau )$]
in the system. It is worth noting that, for a macroscopic system, the $%
\delta $ poles in the spectral representation (\ref{eq:G_SD}) are expected
to be lying infinitesimally close, therewith defining a continuous function $%
\Lambda _{AB}(\omega )$ for real $\omega $. Then one can speculate that the
excitation concept may work only under the basic assumption that the $q$ SD
exhibits some pronounced peaks whose widths have to be considered as a direct
measure of the damping or the life-time of excitations (or $q$%
 quasiparticles, elementary or collective depending on the physical nature
of the operators $A$ and $B$). As mentioned before, this picture should be
associated with the appearance of further complicated singularities of the $q$%
 GF $G_{qAB}(\omega )$ which may occur in the $\omega $ complex plane on the
Riemann sheet below the real axis where $G_{qAB}^{(r)}(\omega )$ is not an
analytical function. Hence, in practical calculations, one must search for
the complex poles of $G_{qAB}^{(r)}(\omega )$ very close to, but below, the
real $\omega$ axis. For each of them, the real part will determine the
frequency of the excitations (the $q$ excitation dispersion relation) and
the imaginary part will represent their damping or life time. In this
scenario, $\Lambda _{qAB}(\omega )$ will result a superposition of
quasi-Lorentian peaks, at characteristic frequencies, whose widths will
represent the damping of the related excitations. Of course, if these widths
reduce to or are zero under appropriate physical conditions, the $q$ SD will
be given by a superposition of $\delta $ functions signaling the
occurrence of undamped excitations in the system.

\subsection{Expressions of the two-time $q$ correlation functions in terms
of the $q$-spectral density\label{sec_3.3}}

In the previous section compact relations between $G_{qAB}^{(\nu )}(\omega )$
and $\Lambda _{qAB}(\omega )$ have been obtained in strict analogy with
the quantum extensive counterpart \cite{Tyablikov:67,Zubarev:74}. We will
show below that the peculiar nature of the Tsallis probability distribution
prevents us from expressing the two-time $q$ CFs $\langle A(\tau )B\rangle _{q}$
and $\langle BA(\tau )\rangle _{q}$ directly in terms of the $q$ SD $\Lambda
_{qAB}(\omega )$ or the related $q$ GFs $G_{qAB}^{(\nu )}(\omega )$ $\left(
\nu =r,a\right) $.

First, it is worth recalling that in extensive quantum statistical
mechanics it is a remarkable feature the existence of direct relations
which allow us to express $\langle A(\tau )B\rangle =\langle A(\tau )B\rangle
_{q=1}$ and $\langle BA(\tau )\rangle =\langle BA(\tau )\rangle _{q=1}$ in
terms of $\Lambda _{AB}(\omega )$ or $G_{AB}^{(\nu )}(\omega )$ $\left( \nu
=r,a\right) $ for two arbitrary operators $A$ and $B$. For future utility we
remember that these relations read \cite{Tyablikov:67,Zubarev:74}
\begin{equation}
\langle A(\tau )B\rangle =\int_{-\infty }^{+\infty }\frac{d\omega }{2\pi }%
\frac{\Lambda _{AB}(\omega )e^{-i\omega \tau }}{1+\eta e^{-\beta \omega }}
\label{eq:Atau_B}
\end{equation}%
and
\begin{equation}
\langle BA(\tau )\rangle =\int_{-\infty }^{+\infty }\frac{d\omega }{2\pi }%
\frac{\Lambda _{AB}(\omega )e^{-i\omega \tau }}{e^{\beta \omega }+\eta },
\label{eq:B_Atau}
\end{equation}%
from which the corresponding static CFs can be immediately obtained setting
$\tau=0$.

The situation becomes sensibly more complicated within the nonextensive
context. To see this in a transparent way, it is convenient to start with
the static $q$ CFs for two arbitrary operators. Taking properly into
account the presence of the $\delta $ functions in the spectral
decomposition (\ref{eq:L_SD}) for the $q$ SD $\Lambda _{qAB}(\omega )$, we
can also write
\begin{multline}
\frac{1}{2\pi }\frac{\Lambda _{qAB}(\omega )}{1+\eta \widetilde{e}_{q}^{%
\hspace{3pt}-\beta \omega }}= \\
=\widetilde{Z}_{q}^{-1}\sum_{n,m}\widetilde{e}%
_{q}^{\hspace{3pt}-\beta \left( E_{n}-U_{q}\right) }
\left\{ \frac{1+\eta \widetilde{e}_{q}^{\hspace{3pt}-\beta \left(
E_{m}-U_{q}\right) }/\widetilde{e}_{q}^{\hspace{3pt}-\beta \left(
E_{n}-U_{q}\right) }}{1+\eta \widetilde{e}_{q}^{\hspace{3pt}-\beta \left(
E_{m}-E_{n}\right) }}\right\} 
A_{nm}B_{mn}\delta \left( \omega -\omega _{mn}\right) ,
\label{eq:q_SD_DF}
\end{multline}%
where we have conveniently introduced the modified $q$ exponential function
\begin{equation}
\widetilde{e}_{q}^{\hspace{3pt}x}=\left[ e_{q}^{x}\right] ^{q}=\left[
1+\left( 1-q\right) x\right] ^{\frac{q}{1-q}}.  \label{eq:m_qef}
\end{equation}%
Then, the integration over $\omega $ easily gives the exact relation
\begin{multline}
\int_{-\infty }^{+\infty }\frac{d\omega }{2\pi }\frac{\Lambda _{qAB}(\omega )%
}{1+\eta \widetilde{e}_{q}^{\hspace{3pt}-\beta \omega }}= \\
= \widetilde{Z}%
_{q}^{-1}\sum_{n,m}\widetilde{e}_{q}^{\hspace{3pt}-\beta \left(
E_{n}-U_{q}\right) }
\left\{ \frac{1+\eta \widetilde{e}_{q}^{\hspace{3pt}-\beta \left(
E_{m}-U_{q}\right) }/\widetilde{e}_{q}^{\hspace{3pt}-\beta \left(
E_{n}-U_{q}\right) }}{1+\eta \widetilde{e}_{q}^{\hspace{3pt}-\beta \left(
E_{m}-E_{n}\right) }}\right\} A_{nm}B_{mn},  \label{eq:q_SD_DF_exrel}
\end{multline}%
with $\widetilde{Z}_{q}=T_{r}\widetilde{e}_{q}^{\hspace{3pt}-\beta \left(
H-U_{q}\right) }$.

By inspection of the right-hand side of this equation, one argues that it
does not reduce to the correlation function $\langle AB\rangle _{q}$ as
happens in the extensive case $q=1$. Nevertheless, if we introduce the
\textquotedblleft $q$ operators\textquotedblright\ $A_{q}$ and $B_{q}$
(related to the original ones $A$ and $B$) defined by the matrix elements
\begin{subequations}
\begin{equation}
A_{qnm}=\langle n|A_{q}|m\rangle =C_{q}\left( n,m\right) A_{nm},
\end{equation}
\begin{equation}
B_{qnm}=\langle n|B_{q}|m\rangle =C_{q}\left( m,n\right) B_{nm},
\end{equation}
\end{subequations}%
where
\begin{equation}
C_{q}\left( n,m\right) =\left\{ \frac{1+\eta \widetilde{e}_{q}^{\hspace{3pt}%
-\beta \left( E_{m}-U_{q}\right) }/\widetilde{e}_{q}^{\hspace{3pt}-\beta
\left( E_{n}-U_{q}\right) }}{1+\eta \widetilde{e}_{q}^{\hspace{3pt}-\beta
\left( E_{m}-E_{n}\right) }}\right\} ^{\frac{1}{2}},  \label{eq:cq}
\end{equation}%
with $C_{q=1}\left( n,m\right) =1$, Eq. (\ref{eq:q_SD_DF_exrel}) yields
\begin{equation}
\langle A_{q}B_{q}\rangle _{q}=\int_{-\infty }^{+\infty }\frac{d\omega }{%
2\pi }\frac{\Lambda _{qAB}(\omega )}{1+\eta \widetilde{e}_{q}^{\hspace{3pt}%
-\beta \omega }}.  \label{eq:aqbq}
\end{equation}%
Similarly, with $\Lambda _{qBA}(\omega )=\eta \Lambda _{qAB}(-\omega )$, one
finds
\begin{equation}
\langle B_{q}A_{q}\rangle _{q}=\int_{-\infty }^{+\infty }\frac{d\omega }{%
2\pi }\frac{\Lambda _{qAB}(\omega )}{\widetilde{e}_{q}^{\hspace{4pt}\beta
\omega }+\eta }.  \label{eq:bqaq}
\end{equation}%
The previous spectral relations, which express the $q$ averages for products
of the two $q$ operators $A_{q}$ and $B_{q}$ in terms of the single $q$ SD $%
\Lambda _{qAB}(\omega )$, are remarkably similar to the extensive static
CFs for $A$ and $B$ and, as expected, they reduce consistently to them for $q\rightarrow 1$.
Of course it is $\langle X_{q}Y_{q}\rangle _{q}\neq \langle
XY\rangle _{q}$ $\left( X,Y=A,B\right) $ for $q\neq 1$.

Concerning the dynamical $q$ CFs for $q$ operators, using Eqs. (\ref%
{eq:F_SD2}), (\ref{eq:q_SD_DF}), and (\ref{eq:cq}) it is now easy to show
that the following relations are true
\begin{equation}
\langle A_{q}(\tau )B_{q}\rangle _{q}=\int_{-\infty }^{+\infty }\frac{%
d\omega }{2\pi }\frac{\Lambda _{qAB}(\omega )e^{-i\omega \tau }}{1+\eta
\widetilde{e}_{q}^{\hspace{3pt}-\beta \omega }}  \label{eq:aqbq3}
\end{equation}%
and
\begin{equation}
\langle B_{q}A_{q}(\tau )\rangle _{q}=\int_{-\infty }^{+\infty }\frac{%
d\omega }{2\pi }\frac{\Lambda _{qAB}(\omega )e^{-i\omega \tau }}{\widetilde{e%
}_{q}^{\hspace{4pt}\beta \omega }+\eta },  \label{eq:bqaq3}
\end{equation}%
which reduce to Eqs. (\ref{eq:Atau_B}) and (\ref{eq:B_Atau}) as $q\rightarrow 1$%
.

The substantial difference with respect to the case $q=1$ lies in the
unfortunate feature that, for $q\neq 1$, the previous compact relations
express the $q$ averages for products of the complicated $q$ operators $%
A_{q} $ and $B_{q}$ in terms of $\Lambda _{qAB}(\omega )$. This may
constitute a serious difficulty in exploring physical cases involving directly
the CFs of the operators $A$ and $B$ which enter the definitions of $%
G_{qAB}^{(\nu )}$ and $\Lambda _{qAB}(\omega )$. Nevertheless, it is
possible to obtain explicit relations, although cumbersome and in general
not very handy, which relate the CFs for the physical operators $A$ and $B$
to those for the corresponding $q$ operators. Indeed, from Eqs. (\ref%
{eq:q_SD_DF_exrel}), (\ref{eq:aqbq}), and (\ref{eq:bqaq}) it is immediate to
show that, for the static case,
\begin{equation}
\langle AB\rangle _{q}=\int_{-\infty }^{+\infty }\frac{d\omega }{2\pi }\frac{%
\Lambda _{qAB}(\omega )}{1+\eta \widetilde{e}_{q}^{\hspace{3pt}-\beta \omega
}}
-\widetilde{Z}_{q}^{-1}\sum_{n,m}\widetilde{e}_{q}^{\hspace{3pt}-\beta
\left( E_{n}-U_{q}\right) }D_{q}(n,m)A_{nm}B_{mn}  \label{eq:abq}
\end{equation}%
and
\begin{equation}
\langle BA\rangle _{q}=\int_{-\infty }^{+\infty }\frac{d\omega }{2\pi }\frac{%
\Lambda _{qAB}(\omega )}{\widetilde{e}_{q}^{\hspace{4pt}\beta \omega }+\eta }%
-\widetilde{Z}_{q}^{-1}\sum_{n,m}\widetilde{e}_{q}^{\hspace{3pt}-\beta
\left( E_{n}-U_{q}\right) }D_{q}(n,m)B_{nm}A_{mn},  \label{eq:baq}
\end{equation}%
with $D_{q}(n,m)=C_{q}^{2}(n,m)-1\rightarrow 0$ as $q\rightarrow 1$. Similar
expressions are true for dynamical CFs $\langle A(\tau )B\rangle _{q}$ and $%
\langle BA(\tau )\rangle _{q}$, which will involve exponential oscillations
in time.

A comparison with the corresponding extensive ones (\ref{eq:Atau_B}) and (\ref%
{eq:B_Atau}) allows us to understand, in a transparent way, the nature of
the deviations from the extensive limit $q=1$. It is worth noting that,
under the condition
\begin{equation}
|\left( 1-q\right) \beta \left( H-U_{q}\right) |\ll 1,  \label{eq:condition}
\end{equation}%
the previous relations simplify to
\begin{equation}
\langle A(\tau )B\rangle _{q}\simeq \int_{-\infty }^{+\infty }\frac{d\omega
}{2\pi }\frac{\Lambda _{qAB}(\omega )e^{-i\omega \tau }}{1+\eta \widetilde{e}%
_{q}^{\hspace{3pt}-\beta \omega }}  \label{eq:ataub_q_approx}
\end{equation}%
and
\begin{equation}
\langle BA(\tau )\rangle _{q}\simeq \int_{-\infty }^{+\infty }\frac{d\omega
}{2\pi }\frac{\Lambda _{qAB}(\omega )e^{-i\omega \tau }}{\widetilde{e}_{q}^{%
\hspace{4pt}\beta \omega }+\eta },  \label{eq:batau_q_approx}
\end{equation}%
which become exact for $q=1$. Of course, the simplified static $q$ CFs for $%
A$ and $B$ can be obtained setting $\tau =0$ in Eqs. (\ref{eq:ataub_q_approx}%
) and (\ref{eq:batau_q_approx}). These equations may be conveniently used in
practical calculations.
Of course, without the restrictive condition (\ref{eq:condition}), one must use the cumbersome Eqs. (\ref{eq:abq}) and (\ref{eq:baq}) and resort to numerical calculations taking properly
into account the cutoff condition (\ref{eq:cond_13}).
It is worth noting that, although the condition (\ref{eq:condition}) is certainly verified in the limit $q\rightarrow1$, it can be also realized for $q$ far from unity with suitable choices of the parameters $\beta$ and $E_{n}-U_{q}$. However, in view of a still reduced number of applications \cite{cavallo:01,Cavallo06}, in order to gain experience in using the $q$ many body formalism developed before for more complex situations, it may be in any case useful to consider weak nonextensivity conditions with $q$ close to unity.

\subsection{Parametric representation for the two-time $q$ Green's functions\label%
{sec_3.4}}

As a conclusion of this section, we will present an alternative way to introduce
the two-time GFs, the related SD, and the two-time CFs by using a
parametric representation, suggested in Refs. \cite{Lenzi:99,Abe:99} in the
context of the Kadanoff-Baym formalism \cite{kadanoff:62}.
This representation allows to
express the relevant $q$ quantities in terms of appropriate parametric
integrals involving the corresponding $q=1$-ones.

The aim is to clarify the statement given in the Introduction about the
effectiveness of the method in practical calculations and to stress again
the potentiality of our framework for a direct calculation of the $q$%
 quantities of interest. The basic idea is to take $b=1-\left( 1-q\right)
\beta \left( H-U_{q}\right) $ and alternatively $z=1+1/(1-q)$, $z=1/(1-q)$,
and $z=q/(1-q)$ in the contour integral representation
\begin{equation}
\Gamma ^{-1}(z)=ib^{1-z}\int_{C}\frac{du}{2\pi }\exp (-ub)(-u)^{-z},
\label{eq:C_IR}
\end{equation}%
with $b>0$ and $\textrm{Re}z>0$. Here $C$ denotes the contour in the $z$ complex
plane which starts form $+\infty $ on the real axis, encircles the origin
once counterclockwise, and returns to $+\infty $. With these ingredients, one
can easily obtain \cite{Lenzi:99,Abe:99} the following representation for $%
Z_{q}(\beta )$, $G_{qAB}^{(\nu )}\left( \tau ;\beta \right) $, and $\langle
A_{q}(\tau )B_{q}\rangle _{q}$ (here we need to explicit the $\beta $%
 dependence)
\begin{equation}
Z_{q}(\beta )=\int_{C}duK_{q}^{(1)}(u)Z_{1}\left[ -\beta u\left( 1-q\right)
\right] ,  \label{eq:Zq_beta}
\end{equation}%
\begin{equation}
G_{qAB}^{(\nu )}(\tau ;\beta ) = \int_{C}duK_{q}^{(2)}(u)Z_{1}\left[ -\beta
u\left( 1-q\right) \right] 
 G_{1AB}^{(\nu )}\left( \tau ;-\beta u\left( 1-q\right) \right) ,
\label{eq:G_qAB_taubeta}
\end{equation}%
and
\begin{equation}
\langle A_{q}(\tau )B_{q}\rangle _{q} = \int_{C}duK_{q}^{(3)}(u)Z_{1}\left[
-\beta u\left( 1-q\right) \right] 
 \langle A(\tau )B\rangle _{1,-\beta u\left( 1-q\right) },
\label{eq:Aqtau_Bq}
\end{equation}%
where
\begin{equation}
K_{q}^{(1)}(u)=\frac{i}{2\pi }\Gamma \left( \frac{2-q}{1-q}\right) e^{-u%
\left[ 1+\left( 1-q\right) \beta U_{q}\right] }(-u)^{-\frac{2-q}{1-q}},
\label{eq:kq1}
\end{equation}%
\begin{equation}
K_{q}^{(2)}(u)=-\frac{\left( 1-q\right) }{(Z_{q})^{q}}K_{q}^{(1)}(u),
\label{eq:kq2}
\end{equation}%
\begin{equation}
K_{q}^{(3)}(u)=\frac{1-q}{q}K_{q}^{(2)}(u),  \label{eq:kq3}
\end{equation}%
and $Z_{1}$, $G_{1AB}^{(\nu )}$, and $\langle A(\tau )B\rangle _{1,\beta }$
denote the corresponding extensive quantities.

An analogous integral representation for the $q$ SD $\Lambda _{qAB}\left(
\omega ;\beta \right) $ can be simply obtained by replacing, in the Fourier
transform of Eq. (\ref{eq:G_qAB_taubeta}), the relation (\ref{eq:qSR}) which
connects $G_{qAB}^{(\nu )}\left( \omega ;\beta \right) $ to $\Lambda
_{qAB}\left( \omega ;\beta \right) $ and, for $q=1$, $G_{1AB}^{(\nu )}\left(
\omega ;\beta \right) $ to the extensive spectral density $\Lambda
_{1AB}\left( \omega ;\beta \right) =\left\langle \left[ A(\tau ),B\right]
_{\eta }\right\rangle _{1,\beta ;\omega }$. One finds
\begin{equation}
\Lambda _{qAB}\left( \omega ;\beta \right)
= \int_{C}duK_{q}^{(2)}Z_{1}\left[ -\beta u\left( 1-q\right) \right]
 \Lambda _{1AB}\left( \omega ;-\beta u\left( 1-q\right) \right) .
\label{eq:lambda_final}
\end{equation}

In view of the previous cumbersome (although elegant) parametric
representation, in our opinion the method to reduce $q$ many-body problems
to the corresponding extensive ones does not appear, in general, convenient
in practical calculations. The two main reasons are (a) it
involves the a priori explicit calculation of ($q=1$)-quantities which,
except for a limited number of simple cases, requires a first step of more
or less reliable approximations; (b) after that, one must
calculate nontrivial contour integrals of complicated functions. In general,
this may be a formidable problem which could require additional
approximations.

Thus, we consider it worthy of interest to develop many-body methods for a
direct study of the $q$ properties overcoming the a priori knowledge of the
related extensive ones. We introduce below the appropriate extensions of two
well-known and powerful quantum many-body methods (for the classical
framework, see Appendix A). Next, we will support our statement by means of
a direct detailed study of the $q$-induced nonextensivity effects on the
low-temperature properties of a nontrivial many-boson model. A preliminary
study, along this direction, of a $d$-dimensional Heisenberg spin model with
long-range interactions was performed in Ref. \cite{Cavallo06}. These are
only two nontrivial many-body problems used by us to test the effectiveness
of our suggestion to perform direct nonextensivity calculations within a
genuine $q$ many-body theory.


\section{Methods for direct calculation of the two-time $q$ Green's functions and
the $q$ spectral density\label{sec_4}}

In this section we present an extension to the nonextensive quantum statistical
mechanics of two intrinsically nonperturbative methods in strict analogy to
the extensive counterpart. These methods will be called the $q$ EMM, for a
direct calculation of the two-time $q$ GFs, and the $q$ SDM, for a direct
calculation of the $q$-SD. In principle, in view of the exact relations
established in the previous section, both the methods are completely
equivalent in the sense that they should give exactly the $q$ GFs and the
related $q$ SD. Nevertheless, previous experiences in quantum \cite%
{kalashnikov:73} and classical \cite{cavallo:05} extensive statistical
mechanics suggest that, in practical calculations, the $q$-SDM may have
several advantages for making more systematic and controllable approximations.

\subsection{$q$ equations-of-motion method\label{sec_4.1}}

By differentiating Eq. (\ref{eq:QNESM2}) with respect $\tau =t-t^{\prime }$
we obtain the $q$ EM for the $q$ GFs in the $\tau $ representation,
\begin{equation}
i\frac{d}{d\tau }\left\langle \left\langle A(\tau );B\right\rangle
\right\rangle _{q}^{(\nu )}=\delta \left( \tau \right) \left\langle \left[
A,B\right] _{\eta }\right\rangle _{q} 
+\left\langle \left\langle \left[ A(\tau ),H\right] _{-};B\right\rangle
\right\rangle _{q}^{(\nu )},  \label{eq:ME1GqAB}
\end{equation}%
or in the $\omega $ representation (more convenient in practical
calculations),
\begin{equation}
\omega \left\langle \left\langle A(\tau);B\right\rangle \right\rangle
_{q,\omega }^{(\nu )}=\left\langle \left[ A,B\right] _{\eta }\right\rangle
_{q}+\left\langle \left\langle \left[ A(\tau),H\right] _{-};B\right\rangle
\right\rangle _{q,\omega }^{(\nu )}.  \label{eq:ME2GqAB}
\end{equation}%
Equations (\ref{eq:ME1GqAB}) and (\ref{eq:ME2GqAB}) are not closed because
higher-order $q$ GFs occur in the problem. Therefore, one needs to consider an
additional EM for these new functions which is again not closed. By
iteration of this procedure, we obtain an infinite hierarchy of coupled EMs
of increasing order which can be written in a compact form as
\begin{equation}
i\frac{d}{d\tau }\left\langle \left\langle L_{H}^{m}A(\tau );B\right\rangle
\right\rangle _{q}^{(\nu )}=\delta \left( \tau \right) \left\langle \left[
L_{H}^{m}A,B\right] _{\eta }\right\rangle _{q}
+\left\langle \left\langle L_{H}^{m+1}A(\tau );B\right\rangle \right\rangle
_{q}^{(\nu )},\text{ \ }(m=0,1,2,...),  \label{eq:EMM1}
\end{equation}%
or
\begin{equation}
\omega \left\langle \left\langle L_{H}^{m}A(\tau);B\right\rangle
\right\rangle _{q,\omega}^{(\nu )}=\left\langle \left[ L_{H}^{m}A,B\right]
_{\eta }\right\rangle _{q} 
+\left\langle \left\langle L_{H}^{m+1}A(\tau);B\right\rangle \right\rangle
_{q,\omega }^{(\nu )}, \quad (m=0,1,2,...),  \label{eq:EMM2}
\end{equation}%
in time and Fourier representation, respectively. Here, we have used the EM (%
\ref{eq:HE}) in the form
\begin{equation}
i\frac{d}{d\tau }A(\tau )=L_{H}^{1}A(\tau )=\left[ A(\tau ),H\right] _{-},
\label{eq:LME}
\end{equation}%
and the operator $L_{H}^{m}$ means $L_{H}^{0}A=A$, $L_{H}^{1}A=\left[ A,H%
\right] _{-}$, $L_{H}^{2}A=\left[ \left[ A,H\right] _{-},H\right] _{-}$, and
so on. Note that the chain of $q$ EMs in the representation (%
\ref{eq:EMM1}) or (\ref{eq:EMM2}) is formally the same for different types
of $q$ GFs and hence one can eliminate the index $\nu $ when the physical
context does not offer ambiguity.

To solve the chain of EMs in the form (\ref{eq:EMM1}) and (\ref{eq:EMM2}),
we must add appropriate boundary conditions which, in the $\omega $%
 representation, can be identified with the asymptotic behaviors (\ref%
{eq:Gq_asymp}). Of course, although Eqs. (\ref{eq:EMM1}) and (\ref{eq:EMM2})
are exact, it is impossible to find a complete solution for interacting
systems. In practical calculations one is forced to use decoupling
procedures, and hence approximate methods, to reduce the infinite chain of
coupled equations to a finite closed one which may be solved. However, in
general, systematic and controllable decouplings are not easy to find and
one must check the reliability of a given approximation by comparing the
results with experiments, simulations or other types of approaches. The $q$%
 SDM, which will be the subject of the next subsection, should be more
flexible in such direction as it happens in the extensive case \cite%
{kalashnikov:73}.

\subsection{$q$-spectral density method\label{sec_4.2}}

By successive derivatives of $\Lambda _{qAB}(\tau )$ [Eq. (\ref{eq:TdqSD})]
with respect to $\tau $ and using the EM (\ref{eq:LME}), we \ have
\begin{equation}
\frac{d^{m}}{d\tau ^{m}}\Lambda _{qAB}(\tau )=(-i)^{m}\left\langle
[L_{H}^{m}A(\tau ),B]_{\eta }\right\rangle _{q}\;(m=0,1,2,...).
\label{eq:SDM_aux}
\end{equation}%
Then, taking the Fourier transform of Eq. (\ref{eq:SDM_aux}) and setting $%
\tau=0$, integration over $\omega $\ yields finally the infinite set of
exact equations for $\Lambda _{qAB}(\omega )$:
\begin{eqnarray}  \label{eq:SDM}
\int_{-\infty }^{+\infty }\frac{d\omega }{2\pi }\omega ^{m}\Lambda
_{qAB}\left( \omega \right) &=&\left\langle [L_{H}^{m}A,B]_{\eta
}\right\rangle _{q}  \notag \\
&=&\left\langle [A,\mathcal{L}_{H}^{m}B]_{\eta }\right\rangle _{q},
\;(m=0,1,2,...),  
\end{eqnarray}%
where the operator $\mathcal{L}_{H}^{m}$ means $\mathcal{L}_{H}^{0}B=B$, $%
\mathcal{L}_{H}^{1}B=\left[ H,B\right] _{-}$, $\mathcal{L}_{H}^{2}B=\left[ H,%
\left[ H,B\right] _{-}\right] _{-}$, and so on. The quantity on the
left-hand side of Eq. (\ref{eq:SDM}) will be called the $m$ moment of
$\Lambda _{qAB}(\omega )$, and the relations (\ref{eq:SDM})
constitute an infinite set of exact moment equations (MEs) or sum rules for
the $q$ SD.

The infinite set (\ref{eq:SDM}) can be seen in a different way. Since the $%
\eta $ commutators and hence the $q$ expectation values involved on the
right-hand side can be calculated, at least in principle, it is quite
remarkable that the $m$ moments of the $q$ SD can be explicitly obtained
without \emph{a priori} knowledge of the function $\Lambda _{qAB}(\omega )$.
This important result implies that the sequence of Eq. (\ref{eq:SDM})
represents a typical moment problem. Its solution would yield the
unknown $q$ SD and hence all the related quantities ($q$ GFs, $q$ CFs and
other observables). Unfortunately, also this problem cannot be solved
exactly and one must look for approximate solutions along the lines
specified below which constitute the key idea of the original SDM \cite%
{kalashnikov:73}.

\subsubsection{Polar ansatz \label{sec_4.2.a}}

As suggested by the exact spectral decomposition (\ref{eq:L_SD}), one seeks
for an approximation of $\Lambda _{qAB}(\omega )$ as a finite sum of
properly weighted $\delta $-functions of the form (\emph{polar ansatz})
\begin{equation}
\Lambda _{qAB}(\omega )=2\pi \sum\limits_{k=1}^{n}\lambda _{qAB}^{(k)}\delta
(\omega -\omega _{qAB}^{(k)}),  \label{eq:PolarAnsatz}
\end{equation}%
where $n$ is an integer number. The unknown parameters $\lambda _{qAB}^{(k)}$
and $\omega _{qAB}^{(k)}$, depending on the physical nature of the operators
$A$ and $B$, have to be determined as a solution of the finite set of $2n$
equations obtained by inserting expression (\ref{eq:PolarAnsatz}) into the
first $2n$ MEs, Eq. (\ref{eq:SDM}). Physically, the parameters $\omega
_{qAB}^{(k)}$ play the role of effective eigenvalues of the Hamiltonian and
each of them represents a real pole of $G_{qAB}(\omega )$ corresponding to a
mode of undamped oscillations for the $q$CF $\left\langle A(\tau
)B\right\rangle _{q}$ [see Eq. (\ref{eq:F_SD1})].

\subsubsection{Modified Gaussian ansatz\label{sec_4.2.b}}

As outlined at the end of Sec. \ref{sec_3.2}, there are physical
situations where the damping of oscillations in the system under study may
be relevant and hence the polar approximation (\ref{eq:PolarAnsatz}) is
inadequate. In these cases, the basic idea of the SDM, related to the moment
problem (\ref{eq:SDM}), remains still valid, but it is necessary to choose a
more appropriate functional structure for the $q$ SD which allows one to
determine the modes of excitations in the system and their damping or
life time. In the extensive context, a generalization of the SDM in this
sense was first proposed by Nolting and Oles \cite{Nolting:80} for Fermi
systems and by Campana et al. \cite{Campana:83} for Bose and classical
systems whose SDs are not positive definite in the whole range of $\omega $%
. Following their key idea, to assure the convergency of the $q$ SD moments
at any order and to preserve the intrinsic physical character of $\Lambda
_{qAB}(\omega )$, one can assume for the $q$ SD the \textit{modified
Gaussian ansatz} \cite{Campana:83}
\begin{equation}
\Lambda _{qAB}(\omega )=2\pi \left( \widetilde{e}_{q}^{\hspace{3pt}\beta
\omega }+\eta \right) \sum\limits_{k=1}^{n}\frac{\lambda _{qAB}^{(k)}}{\sqrt{%
\pi \Gamma _{qAB}^{(k)}}}e^{\frac{-\left( \omega -\omega _{qAB}^{(k)}\right)
^{2}}{\Gamma _{qAB}^{(k)}}}.  \label{eq:GaussianAnsatz}
\end{equation}
Clearly, with the functional representation (\ref{eq:GaussianAnsatz}) for $%
\Lambda _{qAB}(\omega )$, the width of the peak in $\omega =\omega
_{qAB}^{(k)}$ is related to the parameter $\Gamma _{qAB}^{(k)}$ and the
life-time of the excitations with frequency $\omega _{qAB}^{(k)}$ has to be
identified with $\tau _{qAB}^{(k)}=\sqrt{\Gamma _{qAB}^{(k)}}$ under the
condition $\Gamma _{qAB}^{(k)}{/}\left[ \omega _{qAB}^{(k)}\right] ^{2}\ll 1$%
. The choice (\ref{eq:GaussianAnsatz}) is only motivated by the fact that it
makes direct contact with the notation used in the literature for extensive
problems \cite{Nolting:80,Campana:83} and, in view of previous experiences,
it is expected to simplify the algebra also in explicit calculations about $%
q $-induced nonextensivity effects.

As in the $q$ EMM, also in the $q$ SDM the problem remains to close the
truncated finite set of $q$ MEs arising from the polar ansatz (\ref%
{eq:PolarAnsatz}) or the modified Gaussian ansatz (\ref{eq:GaussianAnsatz}).
In any case, evaluation of the right-hand side of Eq. (\ref{eq:SDM})
should generally involve higher-order $q$ SDs. Hence, higher-order moment
problems should be considered and the difficulty of calculations will
increase considerably. So, in order to solve self-consistently the finite
set of $q$ MEs, which arises from Eq. (\ref{eq:SDM}) using the ansatz (\ref%
{eq:PolarAnsatz}) or (\ref{eq:GaussianAnsatz}), it is usually necessary to
use some decoupling procedures and thus to introduce, in a systematic way,
additional consistent approximations in the SDM as in the extensive case
\cite{kalashnikov:73,cavallo:05,Nolting:80,Campana:83}.


\section{Nonextensivity effects for a high-density Bose gas with strong
attraction between particles\label{sec_5}}

\subsection{Model\label{sec_5.1}}

For practical and explicit calculations we consider here a nontrivial Bose
model introduced several years ago by Babichenko \cite{Babichenko:73} with
the aim to explore the properties of a Bose system with strong attraction
between the particles. In contrast with real situations, where the
interparticle interaction potential is characterized by a hard repulsive
core at short distances and a strong attractive well at large distances, in
Ref. \cite{Babichenko:73} the repulsive core was assumed to be soft. This
simplification allowed us to study exactly the properties of the model
at $T=0$, making it possible to use diagrammatic techniques quite similar to
the ones used for high-density Bose systems with a Coulomb pair interaction \cite%
{Brueckner:67}. It is worth noting that, although the choice of a soft
repulsive core does not correspond to the real situation, the possibility to
obtain exact results for such a model is undoubtedly of interest since it
leads to a strongly compressed ground state strictly related to the peculiar
relation between the parameters of the attractive and repulsive parts of the
pair interaction potential. In this sense, the model may be considered as a
complement to the well-studied weakly nonideal Bose gas.

The same high-density Bose model was further studied in Ref. \cite%
{Campana_NC:79} to explore the finite-temperature effects in the
context of the extensive many-body theory. Specifically, the usual
Bogoliubov approximation was proved to be valid for this
model and the exact results by Babichenko \cite{Babichenko:73} were simply
reproduced. Besides, the low-temperature properties of the model were again
achieved without employing cumbersome diagrammatic techniques.

On this ground, also as a further contribution to the general Bose-Einstein
condensation (BEC) scenario, we apply the $q$-many-body methods of Sec. \ref%
{sec_4} to investigate the $q$-induced nonextensivity effects on the
low-temperature behavior of the relevant thermodynamic quantities for the
Babichenko model \cite{Babichenko:73}.

As a first step, we present below the definition of the model. Working in
the grand-canonical ensemble and with periodic boundary conditions, the Bose
model of interest is described by the second-quantized Hamiltonian%
\begin{multline}
\widehat{\mathcal{H}}=\widehat{H}-\mu \widehat{N}=\int d^{3}r\widehat{\psi }%
^{\dagger }(\mathbf{r})\left\{ -\frac{\hslash ^{2}\nabla ^{2}}{2m}-\mu
\right\} \widehat{\psi }(\mathbf{r})+ \\
+\frac{1}{2}\int d^{3}r\int d^{3}r^{\prime }\varphi \left( \left\vert
\mathbf{r}-\mathbf{r}^{\prime }\right\vert \right) \widehat{\psi }^{\dagger
}(\mathbf{r})\widehat{\psi }^{\dagger }(\mathbf{r}^{\prime })\widehat{\psi }(%
\mathbf{r})\widehat{\psi }(\mathbf{r}^{\prime }),
\label{Eq:Babichenko_model}
\end{multline}%
where $\widehat{N}$ is the total number operator of spinless bosons with
mass $m$, $\mu $ is the chemical potential, and $\widehat{\psi }(\mathbf{r})$%
 and $\widehat{\psi }^{\dagger }(\mathbf{r})$ are the usual Bose field
operators. The pair interaction potential $\varphi \left( r\right) $ in (\ref%
{Eq:Babichenko_model}) is assumed \cite{Babichenko:73} as the superposition
of a repulsive Yukawa-like potential of radius $R_{0}$ and of an attractive
Gaussian well of radius $R>R_{0}$ and depth $U_{0}>0$, with the
representation%
\begin{equation}
\varphi \left( r\right) =\frac{\gamma }{r}\exp \left( -\frac{r}{R_{0}}%
\right) -U_{0}\exp \left( -\frac{r^{2}}{R^{2}}\right) .
\label{eq:Babichenko_potential}
\end{equation}%
Here $\gamma $ is a certain definite positive coupling parameter and the
strong attraction or deep-well condition $U_{0}R^{2}\gtrsim \hslash ^{2}/m$
is assumed to be satisfied.

For our purposes it is convenient to choose the system of units with $%
\hslash =m=R_{0}=1$ and to work in the wave-vector $\left\{ \mathbf{k}%
\right\} $ representation. So the grand-canonical Hamiltonian (\ref%
{Eq:Babichenko_model}) assumes the form%
\begin{equation}
\widehat{\mathcal{H}}=\sum_{\mathbf{k}}\varepsilon _{\mathbf{k}}a_{\mathbf{k}%
}^{\dag }a_{\mathbf{k}} 
+\frac{1}{2V}\sum_{\left\{ \mathbf{k}_{\nu }\right\} }\varphi \left(
\left\vert \mathbf{k}_{1}-\mathbf{k}_{3}\right\vert \right) \delta _{\mathbf{%
k}_{1}+\mathbf{k}_{2},\mathbf{k}_{3}+\mathbf{k}_{4}}a_{\mathbf{k}_{1}}^{\dag
}a_{\mathbf{k}_{2}}^{\dag }a_{\mathbf{k}_{3}}a_{\mathbf{k}_{4}},
\label{eq:boson_model}
\end{equation}%
where $\varepsilon _{\mathbf{k}}=k^{2}/2-\mu $, $V$ is the volume of the
system and the Fourier transform $\varphi \left( k\right) $ of the pair
interaction potential (\ref{eq:Babichenko_potential}) is given by%
\begin{equation}
\varphi \left( k\right) =\frac{\gamma }{1+k^{2}}-\pi ^{3/2}U_{0}R^{3}\exp
\left( -\frac{k^{2}R^{2}}{4}\right) ,
\end{equation}%
with%
\begin{equation}
\varphi \left( 0\right) =\gamma -\pi ^{3/2}U_{0}R^{3}=\pi ^{3/2}\eta
U_{0}R^{3},\hspace{15pt}
0<\eta \ll 1,
\end{equation}%
assumed to be small.

With the previous definitions, the coupling parameters involved in the pair
interaction potential are connected by the relation%
\begin{equation}
\gamma =\left( 1+\eta \right) \pi ^{3/2}U_{0}R^{3},
\end{equation}%
and from the strong attraction condition $U_{0}R^{3}\gtrsim 1$, one easily
finds that $\gamma \gtrsim 1$.

It is worth mentioning that, as shown in Ref. \cite{Babichenko:73} and
confirmed by the following calculations, the Gaussian form of the attractive
part of $\varphi \left( r\right) $ is not essential. Indeed, the key
condition to be used through the calculations for the second term in Eq. (%
\ref{eq:Babichenko_potential}) is only a sufficiently smooth change with
distance of $\varphi \left( r\right) $ such that its Fourier transform is
localized in a small region of the wave-vector space ($k\lesssim 1/R$).

\subsection{$q$ equations-of-motion method within the Bogoliubov
approximation\label{sec_5.2}}

Adopting the conventional Bogoliubov approximation,
usually restricted to weak interactions and low
densities \cite{Bogoliubov:47}, but proved to be valid also for high-density
charged Bose gas \cite{Foldy:61} and for the Babichenko high-density Bose
model with strong attraction between the particles \cite{Campana_NC:79}, the
grand-canonical Hamiltonian (\ref{eq:boson_model}) reduces to%
\begin{equation}
\widehat{\mathcal{H}}\simeq -\mu N_{q0}+\frac{1}{2}\frac{N_{q0}^{2}}{V}%
\varphi (0) 
+\sum_{\mathbf{k\neq }0}\left[ f_{q\mathbf{k}}a_{\mathbf{k}}^{\dag }a_{%
\mathbf{k}}+\frac{1}{2}h_{q\mathbf{k}}\left( a_{\mathbf{k}}^{\dag }a_{-%
\mathbf{k}}^{\dag }+a_{\mathbf{k}}a_{-\mathbf{k}}\right) \right] ,
\label{eq:aprx_bose_model}
\end{equation}%
where $N_{q0}$ denotes an unknown $q$ mean number of bosons in the condensate and%
\begin{eqnarray}
f_{q\mathbf{k}} &=&\varepsilon _{\mathbf{k}}+n_{q0}[\varphi (0)+\varphi
\left( \mathbf{k}\right) ], \\
h_{q\mathbf{k}} &=&n_{q0}\varphi \left( \mathbf{k}\right) ,\text{ \ \ \ }%
n_{q0}=\frac{N_{q0}}{V}.
\end{eqnarray}%
According to the standard Bogoliubov picture, one now should diagonalize the
truncated Hamiltonian (\ref{eq:aprx_bose_model}) by a linear canonical
transformation and then proceed, at least in principle, to calculate the OLM
low-temperature $q$-thermodynamic quantities with the prescription outlined
in Sec. \ref{sec_2} (for the extensive case see Ref. \cite{Campana_NC:79}).
However, we find it convenient to follow a different approach which makes
direct contact with the $q$ many-body formalism developed in Secs. \ref%
{sec_3}-\ref{sec_4} and allows us to explore, via direct explicit
calculations, the $q$-induced nonextensivity in the low-temperature regime
in a simpler and more transparent way.

Let us introduce the single-particle retarded and advanced two-time $q$ GFs
\begin{eqnarray}
G_{q\mathbf{k}}^{(\nu )}(\tau ) &=&-i\theta _{\nu }(\tau )\left\langle \left[
a_{\mathbf{k}}(\tau ),a_{\mathbf{k}}^{\dag }\right] _{-}\right\rangle _{q},
\\
\overline{G}_{q\mathbf{k}}^{(\nu )}(\tau ) &=&-i\theta _{\nu }(\tau
)\left\langle \left[ a_{-\mathbf{k}}^{\dag }(\tau ),a_{\mathbf{k}}^{\dag }%
\right] _{-}\right\rangle _{q},
\end{eqnarray}%
where the Heisenberg representation of the operators and the $q$ averages
have to be considered with respect to the truncated Hamiltonian (\ref%
{eq:aprx_bose_model}).

Working in $\omega $ space, for the retarded or advanced $q$ GFs $G_{q%
\mathbf{k}}\left( \omega \right) =\left\langle \left\langle a_{\mathbf{k}%
}(\tau );a_{\mathbf{k}}^{\dag }\right\rangle \right\rangle _{q,\omega }$ and
$\overline{G}_{q\mathbf{k}}\left( \omega \right) =\left\langle \left\langle
a_{-\mathbf{k}}^{\dag }(\tau );a_{\mathbf{k}}^{\dag }\right\rangle
\right\rangle _{q,\omega }$, one easily finds the two coupled EMs
\begin{subequations}
\begin{equation}
\omega G_{q\mathbf{k}}\left( \omega \right) =1+f_{q\mathbf{k}}G_{q\mathbf{k}%
}\left( \omega \right) +h_{q\mathbf{k}}\overline{G}_{q\mathbf{k}}\left(
\omega \right),
\end{equation}
\begin{equation}
\omega \overline{G}_{q\mathbf{k}}\left( \omega \right) =-h_{q\mathbf{k}}G_{q%
\mathbf{k}}\left( \omega \right) -f_{q\mathbf{k}}\overline{G}_{q\mathbf{k}%
}\left( \omega \right) .%
\end{equation}
\end{subequations}%
This algebraic system can be simply solved to find%
\begin{equation}
G_{q\mathbf{k}}\left( \omega \right) =\frac{\omega +f_{q\mathbf{k}}}{\omega
^{2}-\omega _{q\mathbf{k}}^{2}} 
=\frac{1}{2}\left[ \left( 1+\frac{f_{q\mathbf{k}}}{\omega _{q\mathbf{k}}}%
\right) \frac{1}{\omega -\omega _{q\mathbf{k}}}+\left( 1-\frac{f_{q\mathbf{k}%
}}{\omega _{q\mathbf{k}}}\right) \frac{1}{\omega +\omega _{q\mathbf{k}}}%
\right] ,  \label{eq:Gk_boson}
\end{equation}%
\begin{equation}
\overline{G}_{q\mathbf{k}}\left( \omega \right) =\frac{-h_{q\mathbf{k}}}{%
\omega ^{2}-\omega _{q\mathbf{k}}^{2}}=\frac{-h_{q\mathbf{k}}}{2\omega _{q%
\mathbf{k}}}\left( \frac{1}{\omega -\omega _{q\mathbf{k}}}-\frac{1}{\omega
+\omega _{q\mathbf{k}}}\right) .  \label{eq:Gck_boson}
\end{equation}%
Here, the quantity%
\begin{equation}
\omega _{q\mathbf{k}}=\left( f_{q\mathbf{k}}^{2}-h_{q\mathbf{k}}^{2}\right)
^{\frac{1}{2}} 
=\left\{ \left[ \frac{k^{2}}{2}-\mu +n_{q0}\left( \varphi (0)+\varphi
(k)\right) \right] ^{2}-n_{q0}^{2}\varphi ^{2}(k)\right\} ^{\frac{1}{2}},
\end{equation}%
represents the energy (frequency) spectrum of the undamped elementary
excitations in the system. To obtain the relevant $q$ physical quantities it
is now convenient to introduce the single-particle $q$ SDs for the two $q$%
 GFs $G_{q\mathbf{k}}$ and $\overline{G}_{q\mathbf{k}}$ defined as [see
Eqs. (\ref{eq:TdqSD}) and (\ref{eq:TdqSD_FT})]%
\begin{equation}
\Lambda _{q\mathbf{k}}(\omega )=\left\langle \left[ a_{\mathbf{k}}(\tau ),a_{%
\mathbf{k}}^{\dag }\right] _{-}\right\rangle _{q,\omega },
\label{eq:SD1q_boson}
\end{equation}%
and%
\begin{equation}
\overline{\Lambda }_{q\mathbf{k}}(\omega )=\left\langle \left[ a_{-\mathbf{k}%
}^{\dag }(\tau ),a_{\mathbf{k}}^{\dag }\right] _{-}\right\rangle _{q,\omega
},  \label{eq:SD2q_boson}
\end{equation}%
respectively.

Then, from Eqs. (\ref{eq:qSD_omegarep}), (\ref{eq:Gk_boson}), and (\ref%
{eq:Gck_boson}), we immediately have the two $\delta $-function
representations for the spectral densities:%
\begin{equation}
\Lambda _{q\mathbf{k}}(\omega )=\pi \left[ (1+\gamma _{q\mathbf{k}})\delta
\left( \omega -\omega _{q\mathbf{k}}\right) +(1-\gamma _{q\mathbf{k}})\delta
\left( \omega +\omega _{q\mathbf{k}}\right) \right]   \label{eq:2deltSDq}
\end{equation}%
and%
\begin{equation}
\overline{\Lambda }_{q\mathbf{k}}(\omega )=\pi \lambda _{q\mathbf{k}}\left[
\delta \left( \omega +\omega _{q\mathbf{k}}\right) -\delta \left( \omega
-\omega _{q\mathbf{k}}\right) \right] ,  \label{eq:2deltSDq2}
\end{equation}%
where%
\begin{equation}
\gamma _{q\mathbf{k}}=\frac{f_{q\mathbf{k}}}{\omega _{q\mathbf{k}}},\text{ \
\ \ \ }\lambda _{q\mathbf{k}}=\frac{h_{q\mathbf{k}}}{\omega _{q\mathbf{k}}}.
\label{eq:aux1}
\end{equation}%
We are now in the position to formally determine the relevant $q$%
-thermodynamic quantities simply taking into account the general relations
stated in Sec. \ref{sec_3}.

First, we focus on the chemical potential that, within the Bogoliubov
scenario, is defined by \cite{Fetter:03} $\mu =\left\langle \partial
\widehat{H}/\partial N_{q0}\right\rangle _{q}$. Since $\widehat{H}=\widehat{%
\mathcal{H}}+\mu \widehat{N}$, with $\widehat{N}\simeq N_{q0}+\sum_{\mathbf{k%
}\neq 0}a_{\mathbf{k}}^{\dag }a_{\mathbf{k}}$, Eq. (\ref{eq:aprx_bose_model}%
) yields%
\begin{equation}
\widehat{H}\simeq \mu \sum_{\mathbf{k}\neq 0}a_{\mathbf{k}}^{\dag }a_{%
\mathbf{k}}+\frac{1}{2}\frac{N_{q0}^{2}}{V}\varphi (0) 
+\sum_{\mathbf{k\neq }0}\left[ f_{q\mathbf{k}}a_{\mathbf{k}}^{\dag }a_{%
\mathbf{k}}+\frac{1}{2}h_{q\mathbf{k}}\left( a_{\mathbf{k}}^{\dag }a_{-%
\mathbf{k}}^{\dag }+a_{\mathbf{k}}a_{-\mathbf{k}}\right) \right] .
\end{equation}%
Then, with simple algebra, we get%
\begin{equation}
\mu =n_{q0}\varphi (0) 
+\frac{1}{n_{q0}V}\sum_{\mathbf{k\neq }0}\left[ \left( f_{q\mathbf{k}%
}-\varepsilon _{\mathbf{k}}\right) \left\langle a_{\mathbf{k}}^{\dag }a_{%
\mathbf{k}}\right\rangle _{q}+h_{q\mathbf{k}}\left\langle a_{\mathbf{k}%
}^{\dag }a_{-\mathbf{k}}^{\dag }\right\rangle _{q}\right] .
\label{eq:ChemPot}
\end{equation}%
Besides, for the $q$ internal energy $U_{q}=\left\langle \widehat{H}%
\right\rangle _{q}$, we find%
\begin{equation}
U_{q}=\mu \left( N-N_{q0}\right) +\frac{1}{2}\frac{N_{q0}^{2}}{V}\varphi (0)
+\sum_{\mathbf{k\neq }0}\left[ f_{q\mathbf{k}}\left\langle a_{\mathbf{k}%
}^{\dag }a_{\mathbf{k}}\right\rangle _{q}+h_{q\mathbf{k}}\left\langle a_{%
\mathbf{k}}^{\dag }a_{-\mathbf{k}}^{\dag }\right\rangle _{q}\right] .
\label{eq:Uq_boson}
\end{equation}%
In Eq. (\ref{eq:Uq_boson}), $N=\left\langle \widehat{N}\right\rangle _{q}$
and $N-N_{q0}=\sum_{\mathbf{k\neq }0}\left\langle a_{\mathbf{k}}^{\dag }a_{%
\mathbf{k}}\right\rangle _{q}$, where $N$ is the total number of bosons
in the system. In particular, the $q$ depletion of
condensate will be given by%
\begin{equation}
n-n_{q0}=\frac{1}{V}\sum_{\mathbf{k\neq }0}\left\langle a_{\mathbf{k}}^{\dag
}a_{\mathbf{k}}\right\rangle _{q},  \label{eq:Depleption}
\end{equation}%
where $n=N/V$ is the total number density of particles. Notice that, using Eq. (%
\ref{eq:Depleption}), expressions (\ref{eq:ChemPot}) and (\ref%
{eq:Uq_boson}) for $\mu $ and $\left\langle \widehat{H}\right\rangle _{q}$
can be also written in the most compact form%
\begin{equation}
\mu =n\varphi (0)+\frac{1}{V}\sum_{\mathbf{k\neq }0}\varphi (k)\left[
\left\langle a_{\mathbf{k}}^{\dag }a_{\mathbf{k}}\right\rangle
_{q}+\left\langle a_{\mathbf{k}}^{\dag }a_{-\mathbf{k}}^{\dag }\right\rangle
_{q}\right] ,
\end{equation}%
and%
\begin{equation}
\left\langle \widehat{N}\right\rangle _{q}=N_{q0}(\mu -n_{q0}\varphi (0))+%
\frac{1}{2}\frac{N_{q0}^{2}\varphi (0)}{V}+\frac{1}{2}\sum_{\mathbf{k\neq }%
0}k^{2}\left\langle a_{\mathbf{k}}^{\dag }a_{\mathbf{k}}\right\rangle _{q}.
\end{equation}

Due to the complicated relations (\ref{eq:abq}) and (\ref{eq:baq}), it is not
easy, in general, to calculate explicitly the $q$ averages which enter
Eqs. (\ref{eq:ChemPot})-(\ref{eq:Depleption}) in terms of the $q$ SDs
$\Lambda _{q\mathbf{k}}(\omega )$ and $\overline{\Lambda }_{q\mathbf{k}%
}(\omega )$. Intricate numerical calculations become necessary, or one
must resort to reliable approximations to obtain analytical results.
The problem becomes sensibly handier under the condition [see Eq. (\ref{eq:condition})]
\begin{equation}
\left\vert \left( 1-q\right) \beta \left( \widehat{\mathcal{H}}-\mathcal{U}%
_{q}\right) \right\vert \ll 1,
\end{equation}%
with $\mathcal{U}_{q}=U_{q}-\mu \left\langle \widehat{N}\right\rangle _{q}$.
Under this condition, from Eq. (\ref{eq:ataub_q_approx}) it
immediately follows that
\begin{equation}
\left\langle a_{\mathbf{k}}^{\dag }a_{\mathbf{k}}\right\rangle
_{q}=\int_{-\infty }^{+\infty }\frac{d\omega }{2\pi }\frac{\Lambda _{q%
\mathbf{k}}(\omega )}{\widetilde{e}_{q}^{\hspace{3pt}\beta \omega }-1}
\label{eq:Av1q}
\end{equation}%
and%
\begin{equation}
\left\langle a_{\mathbf{k}}^{\dag }a_{-\mathbf{k}}^{\dag }\right\rangle
_{q}=\int_{-\infty }^{+\infty }\frac{d\omega }{2\pi }\frac{\overline{\Lambda
}_{q\mathbf{k}}(\omega )}{\widetilde{e}_{q}^{\hspace{3pt}\beta \omega }-1}.
\label{eq:Av2q}
\end{equation}%
On the other hand, in the present Bogoliubov scenario, $\Lambda _{q\mathbf{k}%
}(\omega )$ and $\overline{\Lambda }_{q\mathbf{k}}(\omega )$ are exactly
expressed by Eqs. (\ref{eq:2deltSDq}) and (\ref{eq:2deltSDq2}). Then, from Eqs.
(\ref{eq:Av1q}) and (\ref{eq:Av2q}) we easily find%
\begin{equation}
\left\langle a_{\mathbf{k}}^{\dag }a_{\mathbf{k}}\right\rangle _{q}=\frac{1}{%
2}\left\{ \frac{1+\gamma _{q\mathbf{k}}}{\widetilde{e}_{q}^{\hspace{3pt}%
\beta \omega _{q\mathbf{k}}}-1}+\frac{1-\gamma _{q\mathbf{k}}}{\widetilde{e}%
_{q}^{\hspace{3pt}-\beta \omega _{q\mathbf{k}}}-1}\right\}
\label{eq:Nk_bose}
\end{equation}%
and%
\begin{equation}
\left\langle a_{\mathbf{k}}^{\dag }a_{-\mathbf{k}}^{\dag }\right\rangle _{q}=%
\frac{1}{2}\lambda _{q\mathbf{k}}\left\{ {\frac{1}{\widetilde{e}_{q}^{%
\hspace{3pt}-\beta \omega _{q\mathbf{k}}}-1}-\frac{1}{\widetilde{e}_{q}^{%
\hspace{3pt}\beta \omega _{q\mathbf{k}}}-1}}\right\} ,
\end{equation}%
where $\gamma _{q\mathbf{k}}$ and $\lambda _{q\mathbf{k}}$ are given by Eqs.
(\ref{eq:aux1}).

To calculate explicitly and consistently the $q$-thermodynamic quantities as
functions of the temperature $T$ and the total number density $n$, one must
now express the chemical potential as a function of $T$ and $n$, performing
the sums over $\mathbf{k}$ in the previous basic expressions. This is not an
easy task since, as $q$, $\mu $ enters the problem in a cumbersome way through the
energy spectrum $\omega _{q\mathbf{k}}$. However, it will be shown below
that, in the physical regimes of interest (high-density and
low-temperature regimes), one has $\left\vert \mu -\mu _{q0}\right\vert /\mu_{q0} \ll 1$,
with $\mu _{q0}=n_{q0}\varphi (0)$, and, consistently, $\left(
n-n_{q0}\right) /n=\left( N-N_{q0}\right) /N\ll 1$, as expected in a
Bogoliubov approximation scenario. This key feature allows us to simplify
sensibly the problem setting, as a good (first) approximation, $\mu \simeq
\mu _{q0}=n_{q0}\varphi (0)=(n_{q0}\gamma )\eta /(1+\eta )$ and also $%
n_{q0}\simeq n$ in the expressions of $\varepsilon _{\mathbf{k}}$, $\omega
_{q\mathbf{k}}$, and $f_{q\mathbf{k}}$, which enter the previous general
relations for the relevant $q$-thermodynamic quantities. So, in particular,
with $\mu \simeq \mu _{q0}+\Delta \mu $, the correction $\Delta \mu $ to the
chemical potential will be given by the last two terms of Eq. (\ref%
{eq:ChemPot}) with $\mu $\ replaced by $n\varphi (0)$. In all cases, for
explicit calculations, the expressions of $\varepsilon _{\mathbf{k}}$, $%
\omega _{q\mathbf{k}}$ and $f_{q\mathbf{k}}$ in the summands can be
approximated, to leading order, as%
\begin{eqnarray}
\varepsilon _{\mathbf{k}} &\approx &\frac{k^{2}}{2}-n\varphi (0),
\label{eq:eps1k} \\
\omega _{q\mathbf{k}} &\approx &\left[ \frac{k^{4}}{4}+nk^{2}\varphi (%
\mathbf{k})\right] ^{\frac{1}{2}}\equiv \omega _{\mathbf{k}},
\label{eq:omg1k} \\
f_{q\mathbf{k}} &\approx &\frac{k^{2}}{2}+n\varphi (\mathbf{k})\equiv f_{%
\mathbf{k}}.  \label{eq:f1k}
\end{eqnarray}%
Notice that $\omega _{\mathbf{k}}$, on the right-hand side of expression
(\ref{eq:omg1k}), is just the Bogoliubov energy spectrum of elementary
excitations.

After that, with $V^{-1}\sum_{\mathbf{k}\neq 0}(...)\overset{V\rightarrow
\infty }{\longrightarrow }(2\pi ^{2})^{-1}\int_{0}^{\infty }dkk^{2}(...)$,
very complicated integrals remain again to be performed
due to the presence of the power-law function $\widetilde{e}_{q}^{\hspace{3pt}x}$
involved in the integrands. Since resorting to numerical calculations by variation
of $\beta$ and $q$ is beyond the purposes of the present work, to obtain demonstrative explicit results one
is then compelled to make further suitable approximations as the known factorization
procedure \cite{xyx} or expansions close to the extensive value $q=1$ of the
Tsallis parameter. We follow here the second direction
because it is simple and instructive to estimate, to the leading order in $q-1$%
, the $q$ nonextensive effects on the already known exact extensive results
\cite{Campana_NC:79} in the high-density and low-temperature regimes where
the integrals can be explicitly calculated.

To the first order in $q-1$ one obtains the expansion%
\begin{equation}
\frac{1}{\widetilde{e}_{q}^{\hspace{3pt}\pm \beta \omega _{\mathbf{k}}}-1}%
\simeq \frac{1}{e^{\pm \beta \omega _{\mathbf{k}}}-1} 
+\frac{1}{2}(1-q)\frac{e^{\pm \beta \omega _{\mathbf{k}}}}{\left( e^{\pm
\beta \omega _{\mathbf{k}}}-1\right) ^{2}}\left[ \beta ^{2}\omega _{\mathbf{k%
}}^{2}\pm 2\beta \omega _{\mathbf{k}}\right] .
\end{equation}%
So, for the basic $q$-thermodynamic quantities, we can formally write (with
a cumbersome but simple algebra)%
\begin{multline}
n-n_{q0}\simeq \frac{1}{2V}\sum_{\mathbf{k}\neq 0}\frac{f_{\mathbf{k}%
}-\omega _{\mathbf{k}}}{\omega _{\mathbf{k}}}+\frac{1}{V}\sum_{\mathbf{k}%
\neq 0}\frac{f_{\mathbf{k}}}{\omega _{\mathbf{k}}}\frac{1}{e^{\beta \omega _{%
\mathbf{k}}}-1}+ \\
+\frac{1}{2}(1-q)\left\{ \frac{\beta ^{2}}{V}\sum_{\mathbf{k}\neq 0}\frac{%
\omega _{\mathbf{k}}^{2}e^{\beta \omega _{\mathbf{k}}}}{\left( e^{\beta
\omega _{\mathbf{k}}}-1\right) ^{2}}+2\frac{\beta }{V}\sum_{\mathbf{k}\neq 0}%
\frac{f_{\mathbf{k}}e^{\beta \omega _{\mathbf{k}}}}{\left( e^{\beta \omega _{%
\mathbf{k}}}-1\right) ^{2}}\right\} ,  \label{eq:DEPq}
\end{multline}%
\begin{multline}
\mu \simeq n\varphi (0)+\frac{1}{n}\left[ \frac{-1}{2V}\sum_{\mathbf{k}\neq
0}\left( f_{\mathbf{k}}-\omega _{\mathbf{k}}\right) +\frac{1}{V}\sum_{%
\mathbf{k}\neq 0}\frac{\omega _{\mathbf{k}}}{e^{\beta \omega _{\mathbf{k}}}-1%
}\right] + \\
-\frac{1}{2n}\left[ \frac{1}{2V}\sum_{\mathbf{k}\neq 0}k^{2}\frac{f_{\mathbf{%
k}}-\omega _{\mathbf{k}}}{\omega _{\mathbf{k}}}+\frac{1}{V}\sum_{\mathbf{k}%
\neq 0}\frac{k^{2}f_{\mathbf{k}}}{\omega _{\mathbf{k}}}\frac{1}{e^{\beta
\omega _{\mathbf{k}}}-1}\right] + \\
+\frac{1}{2}(1-q)\left\{ \frac{\beta ^{2}}{V}\sum_{\mathbf{k}\neq 0}\frac{%
\omega _{\mathbf{k}}^{2}\varphi \left( k\right) e^{\beta \omega _{\mathbf{k}%
}}}{\left( e^{\beta \omega _{\mathbf{k}}}-1\right) ^{2}}+2\frac{\beta }{V}%
\sum_{\mathbf{k}\neq 0}\frac{k^{2}\varphi \left( k\right) e^{\beta \omega _{%
\mathbf{k}}}}{\left( e^{\beta \omega _{\mathbf{k}}}-1\right) ^{2}}\right\} ,
\label{eq: MUq}
\end{multline}%
\begin{multline}
\left\langle \widehat{H}\right\rangle _{q}\simeq \frac{1}{2}\frac{N^{2}}{V}%
\varphi \left( 0\right) -\frac{1}{2}\sum_{\mathbf{k}\neq 0}\left( f_{\mathbf{%
k}}-\omega _{\mathbf{k}}\right) +\sum_{\mathbf{k}\neq 0}\frac{\omega _{%
\mathbf{k}}}{e^{\beta \omega _{\mathbf{k}}}-1}+ \\
+\frac{1}{2}(1-q)\left\{ \beta ^{2}\sum_{\mathbf{k}\neq 0}\frac{\omega _{%
\mathbf{k}}^{2}f_{\mathbf{k}}e^{\beta \omega _{\mathbf{k}}}}{\left( e^{\beta
\omega _{\mathbf{k}}}-1\right) ^{2}}+2\beta \sum_{\mathbf{k}\neq 0}\frac{%
\omega _{\mathbf{k}}^{2}e^{\beta \omega _{\mathbf{k}}}}{\left( e^{\beta
\omega _{\mathbf{k}}}-1\right) ^{2}}\right\} .  \label{eq:ENERq}
\end{multline}

Now all the integrals in Eqs. (\ref{eq:DEPq})-(\ref{eq:ENERq}) can be
consistently calculated under the conditions $\left( n\gamma \right) ^{1/2}\gg 1$
(high-density regime) and $\beta \gg 1$ (low-temperature regime) where it is
reasonable to speculate that the Bogoliubov approximation preserves its
validity \cite{Campana_NC:79}. Indeed, making the transformation $%
x=k(n\gamma )^{1/4}$, one sees that the principal contribution to the
integrals is made by $x\sim 1$ so that neglecting $O(\left( n\gamma
\right) ^{-1/2})$ or exponentially small terms is quite legitimate \cite%
{Babichenko:73}. A similar situation arises also in a high-density Bose gas with
Coulomb interaction between bosons \cite{Brueckner:67,Foldy:61}. Bearing
this in mind, in the high-density and low-temperature limits, for the
depletion of condensate, the chemical potential, and the $q$ internal energy
density $u_{q}=U_{q}/V=\left\langle \widehat{H}\right\rangle _{q}/V$ as
functions of $T$ and $n$, one finds (see Appendix B for calculations of the
integrals)%
\begin{multline}
n-n_{q0}\simeq a\left( n\gamma \right) ^{\frac{3}{4}}+\frac{1}{12}\left(
\frac{\eta +1}{\eta }\right) ^{\frac{1}{2}}\left( n\gamma \right) ^{-\frac{1%
}{2}}T^{2}+ \\
+\left( 1-q\right) \frac{1}{6}\left( \frac{\eta +1}{\eta }\right) ^{\frac{1}{%
2}}\left( n\gamma \right) ^{-\frac{1}{2}} 
 T^{2}\left\{ 1+\frac{2\pi ^{2}}{5}\left( \frac{\eta +1}{\eta }\right)
\left( n\gamma \right) ^{-1}T\right\} ,  \label{eq:DEP1q}
\end{multline}%
\begin{multline}
\mu \simeq \frac{\eta +1}{\eta }\left( n\gamma \right) -b\gamma \left(
n\gamma \right) ^{\frac{1}{4}} 
+\frac{\pi ^{2}}{60}\left( \frac{\eta +1}{\eta }\right) ^{\frac{3}{2}}\gamma
\left( n\gamma \right) ^{-\frac{5}{2}}T^{4}+ \\
+\left( 1-q\right) \frac{\pi ^{2}}{15}\left( \frac{\eta +1}{\eta }\right) ^{%
\frac{1}{2}}\gamma \left( n\gamma \right) ^{-\frac{3}{2}} 
 T^{3}\left\{ 1+2\left( \frac{\eta +1}{\eta }\right) \left( n\gamma
\right) ^{-1}T\right\} ,  \label{eq:MU1q}
\end{multline}%
\begin{multline}
u_{q}\simeq \frac{1}{2}\frac{\eta +1}{\eta }\frac{1}{\gamma }\left( n\gamma
\right) ^{2}-\frac{4}{5}b\left( n\gamma \right) ^{\frac{5}{4}} 
+\frac{\pi ^{2}}{30}\left( \frac{\eta +1}{\eta }\right) ^{\frac{3}{2}}\left(
n\gamma \right) ^{-\frac{3}{2}}T^{4}+ \\
+\left( 1-q\right) \frac{\pi ^{2}}{15}\left( \frac{\eta +1}{\eta }\right) ^{%
\frac{1}{2}}\left( n\gamma \right) ^{-\frac{1}{2}} 
 T^{3}\left\{ 1+2\left( \frac{\eta +1}{\eta }\right) \left( n\gamma
\right) ^{-1}T\right\} ,  \label{eq:ENER1q}
\end{multline}%
where $a=\sqrt{2}\Gamma ^{2}(1/4)/(48\pi ^{5/2})$ and $b=\sqrt{2}\Gamma
^{2}(3/4)/(4\pi ^{5/2})$.

Notice that, from Eqs. (\ref{eq:DEP1q}) and (\ref{eq:ENER1q}), for $\left(
n_{q0}\gamma \right) ^{1/2}\simeq \left( n\gamma \right) ^{1/2}\gg 1$ and $%
T\ll 1$ one has $\left( n-n_{q0}\right) /n\ll 1$ and $\left\vert \mu -\mu
_{q0}\right\vert /\mu _{q0}\ll 1$, as expected, signaling the internal
consistency of calculations. Of course, for $q=1$ the extensive results are
exactly reproduced \cite{Babichenko:73,Campana_NC:79}. From the previous
basic relations one can now calculate all the $q$-thermodynamic quantities
of the system using the OLM formalism \cite{martinez:00,Ferri:05}. For
instance, the $q$ specific heat at constant volume is given by%
\begin{multline}
C_{qV}=\left( \frac{\partial u_{q}}{\partial T}\right) _{V}\simeq \frac{2\pi
^{2}}{15}\left( \frac{\eta +1}{\eta }\right) ^{\frac{3}{2}}\left( n\gamma
\right) ^{-\frac{3}{2}}T^{3}+ \\
+\left( 1-q\right) \frac{\pi ^{2}}{5}\left( \frac{\eta +1}{\eta }\right) ^{%
\frac{1}{2}}\left( n\gamma \right) ^{-\frac{1}{2}} 
 T^{2}\left\{ 1+\frac{8}{3}\left( \frac{\eta +1}{\eta }\right) \left(
n\gamma \right) ^{-1}T\right\} .  \label{eq:Cvq}
\end{multline}%
It is worth noting that, as suggested by previous results, the $q$-induced
thermal effects may compete sensibly with the extensive contributions.

\subsection{$q$-spectral density method at work: The two-pole
approximation\label{sec_5.3}}

In this section we apply the $q$ SDM directly to the grand-canonical
Hamiltonian (\ref{eq:boson_model}) by avoiding the Bogoliubov approximation
and hence the known troubles related to a truncated Hamiltonian which no
longer conserves the particle number and to the unwanted dependence of $%
N_{q0}$.

Taking advantage from the previous results, we will work rather via a two-$%
\delta $-function ansatz for the appropriate $q$ SD as suggested by Eq. (%
\ref{eq:2deltSDq}). Then, we will establish the conditions for which the results
of Sec. \ref{sec_5.2} can be reproduced in the high-density and
low-temperature regimes. As we shall see, the procedure will also open a
window towards a possible systematic study of the $q$-thermodynamics of
other more realistic second-quantized many-boson models.

For the Hamiltonian (\ref{eq:boson_model}) in the wave-vector
representation, we introduce the single-particle $q$ SD%
\begin{equation}
\Lambda _{q\mathbf{k}}\left( \omega \right) =\left\langle \left[ a_{\mathbf{k%
}}(\tau ),a_{\mathbf{k}}^{\dagger }\right] _{-}\right\rangle _{q,\omega }.
\label{eq:qMomentSD}
\end{equation}%
Then, the general $q$ MEs, Eq. (\ref{eq:SDM}), for the present problem can be
written as%
\begin{eqnarray}
\int_{-\infty }^{+\infty }\frac{d\omega }{2\pi }\omega ^{m}\Lambda _{q%
\mathbf{k}}\left( \omega \right) &=&\left\langle [L_{\widehat{\mathcal{H}}%
}^{m}a_{\mathbf{k}},a_{\mathbf{k}}^{\dagger }]_{-}\right\rangle _{q}  \notag
\\
&=&\left\langle [a_{\mathbf{k}},\mathcal{L}_{\widehat{\mathcal{H}}}^{m}a_{%
\mathbf{k}}^{\dagger }]_{-}\right\rangle _{q},\quad (m=0,1,2,..).  
\end{eqnarray}%
Using the usual bosonic canonical commutation relations, tedious but simple
algebra yields for $m=0,1,2$ (the case of interest for us)
\begin{equation}
\int_{-\infty }^{+\infty }\frac{d\omega }{2\pi }\Lambda _{q\mathbf{k}}\left(
\omega \right) =1,  \label{eq:MEboson1}
\end{equation}%
\begin{equation}
\int_{-\infty }^{+\infty }\frac{d\omega }{2\pi }\omega \Lambda _{q\mathbf{k}%
}\left( \omega \right) =\varepsilon _{\mathbf{k}}+\varphi (0)n 
+ \frac{1}{V}\sum_{\mathbf{k}^{\prime }}\varphi (\left\vert \mathbf{k}-\mathbf{%
k}^{\prime }\right\vert )N_{q\mathbf{k}^{\prime }},  \label{eq:MEboson2}
\end{equation}%
\begin{equation}
\int_{-\infty }^{+\infty }\frac{d\omega }{2\pi }\omega ^{2}\Lambda _{q%
\mathbf{k}}\left( \omega \right) =-\varepsilon _{\mathbf{k}%
}^{2}+2\varepsilon _{\mathbf{k}}\int_{-\infty }^{+\infty }\frac{d\omega }{%
2\pi }\omega \Lambda _{q\mathbf{k}}\left( \omega \right) 
+ L_{q}+\frac{2}{V}\sum_{\mathbf{k}^{\prime }}\varphi (\left\vert \mathbf{k}-%
\mathbf{k}^{\prime }\right\vert )\lambda _{q\mathbf{k}^{\prime }},
\label{eq:MEboson3}
\end{equation}%
where
\begin{eqnarray}
n &=&\frac{1}{V}\sum_{\mathbf{k}}\left\langle a_{\mathbf{k}}^{\dagger }a_{%
\mathbf{k}}\right\rangle _{q}, \\
N_{q\mathbf{k}} &=&\left\langle a_{\mathbf{k}}^{\dagger }a_{\mathbf{k}%
}\right\rangle _{q},  \label{eq:122} \\
L_{q} &=&\frac{1}{V^{2}}\sum_{\left\{ \mathbf{k}_{\nu }\right\} }\delta _{%
\mathbf{k}_{1}+\mathbf{k}_{2};\mathbf{k}_{3}+\mathbf{k}_{4}}\varphi
^{2}\left( \left\vert \mathbf{k}_{1}-\mathbf{k}_{3}\right\vert \right)
 \left\langle a_{\mathbf{k}_{1}}^{\dagger }a_{\mathbf{k}_{3}}a_{%
\mathbf{k}_{2}}^{\dagger }a_{\mathbf{k}_{4}}\right\rangle _{q},
\end{eqnarray}%
and%
\begin{equation}
\lambda _{q\mathbf{k}}=\frac{1}{2V}\sum_{\mathbf{k}^{\prime }}\varphi
(\left\vert \mathbf{k}-\mathbf{k}^{\prime }\right\vert )N_{q\mathbf{k}%
^{\prime }} 
+\frac{1}{V}\sum_{\left\{ \mathbf{k}_{1},\mathbf{k}_{2},\mathbf{k}%
_{4}\right\} }\varphi \left( \left\vert \mathbf{k}_{1}-\mathbf{k}\right\vert
\right) \delta _{\mathbf{k}_{1}+\mathbf{k}_{2};\mathbf{k}+\mathbf{k}%
_{4}}\left\langle a_{\mathbf{k}_{1}}^{\dagger }a_{\mathbf{k}_{2}}^{\dagger
}a_{\mathbf{k}}a_{\mathbf{k}_{4}}\right\rangle _{q}.  \label{eq:aux132}
\end{equation}%
As we see, $L_{q}$ and $\lambda _{q\mathbf{k}}$ introduce in the problem
two-particle $q$ CFs which should be expressed in terms of $\Lambda _{q%
\mathbf{k}}\left( \omega \right) $ to close the truncated system of MEs (%
\ref{eq:MEboson1})-(\ref{eq:MEboson3}). Unfortunately, this constitutes a
serious difficulty since, as shown in Sec. \ref{sec_3}, there is no
simple relation between a CF of the type $\left\langle BA\right\rangle _{q}$
and the corresponding $q$ SD $\Lambda _{qAB}\left( \omega \right) $.
However, under the condition $\left\vert \left( 1-q\right) \beta \left( \widehat{%
\mathcal{H}}-\mathcal{U}_{q}\right) \right\vert \ll 1$, the
calculations simplify sensibly. Indeed, in this case, one can assume $%
\left\langle BA\right\rangle _{q}\simeq $ $\left( 2\pi \right)
^{-1}\int_{-\infty }^{+\infty }d\omega \Lambda _{qAB}\left( \omega \right) /(%
\widetilde{e}_{q}^{\hspace{3pt}\beta \omega }-1)$ and in Eqs. (\ref%
{eq:MEboson2})-(\ref{eq:122}) and (\ref{eq:aux132}) one has%
\begin{equation}
N_{q\mathbf{k}}=\int_{-\infty }^{+\infty }\frac{d\omega }{2\pi }\frac{%
\Lambda _{q\mathbf{k}}\left( \omega \right) }{\widetilde{e}_{q}^{\hspace{3pt}%
\beta \omega }-1}.  \label{eq:nksegnato}
\end{equation}%
Besides, with standard straightforward algebra \cite%
{kalashnikov:73,Campana_NC:79}, we have also
\begin{equation}
\frac{1}{V}\sum_{\left\{ \mathbf{k}_{1},\mathbf{k}_{2},\mathbf{k}%
_{4}\right\} }\varphi (\left\vert \mathbf{k}_{1}-\mathbf{k}\right\vert
)\delta _{\mathbf{k}_{1}+\mathbf{k}_{2};\mathbf{k}+\mathbf{k}%
_{4}}\left\langle a_{\mathbf{k}_{1}}^{\dagger }a_{\mathbf{k}_{2}}^{\dagger
}a_{\mathbf{k}}a_{\mathbf{k}_{4}}\right\rangle _{q} 
=\int_{-\infty }^{+\infty }\frac{d\omega }{2\pi }\frac{\omega -\varepsilon _{%
\mathbf{k}}}{\widetilde{e}_{q}^{\hspace{3pt}\beta \omega }-1}\Lambda _{q%
\mathbf{k}}\left( \omega \right) .
\end{equation}%
This allows us to express $\lambda _{q\mathbf{k}}$ in Eq. (\ref{eq:aux132})
in terms of $\Lambda _{q\mathbf{k}}\left( \omega \right) $ as
\begin{equation}
\lambda _{q\mathbf{k}}=\frac{1}{2V}\sum_{\mathbf{k}^{\prime }}\varphi
(\left\vert \mathbf{k}-\mathbf{k}^{\prime }\right\vert )\int_{-\infty
}^{+\infty }\frac{d\omega }{2\pi }\frac{\Lambda _{q\mathbf{k}^{\prime
}}\left( \omega \right) }{\widetilde{e}_{q}^{\hspace{3pt}\beta \omega }-1}+
 \int_{-\infty }^{+\infty }\frac{d\omega }{2\pi }\frac{\omega -\varepsilon _{%
\mathbf{k}}}{\widetilde{e}_{q}^{\hspace{3pt}\beta \omega }-1}\Lambda _{q%
\mathbf{k}}\left( \omega \right)   \label{eq:lambdakmin}
\end{equation}%
and to obtain for the $q$ internal energy the following expression in terms
of the single-particle $q$ SD:%
\begin{equation}
U_{q}=\left\langle \widehat{H}\right\rangle _{q}=\mu N+\frac{1}{2}\sum_{%
\mathbf{k}}\int_{-\infty }^{+\infty }\frac{d\omega }{2\pi }\frac{\omega
+\varepsilon _{\mathbf{k}}}{\widetilde{e}_{q}^{\hspace{3pt}\beta \omega }-1}%
\Lambda _{q\mathbf{k}}\left( \omega \right) ,
\label{Eq:129}
\end{equation}%
where $N=\sum_{\mathbf{k}}N_{q\mathbf{k}}$.

Nevertheless, the quantity $L_{q}$ cannot be expressed exactly in terms of $%
\Lambda _{q\mathbf{k}}\left( \omega \right) $, so that, the truncated system
(\ref{eq:MEboson1})-(\ref{eq:MEboson3}) is not yet closed and one must
resort to additional decoupling procedures \cite{kalashnikov:73} which allow one
to express also $L_{q}$ in terms of $\Lambda _{q\mathbf{k}}\left( \omega
\right) $.
We shall see below that this problem can be simply solved on physical grounds
in the regime of interest.

Working within
the spirit of the SDM \cite{kalashnikov:73,Campana_NC:79} (see Sec. III),
we assume for $\Lambda_{q\mathbf{k}}\left( \omega \right) $
the two-$\delta$-function ansatz [also suggested by Eq. (\ref{eq:2deltSDq})]
\begin{equation}
\Lambda _{q\mathbf{k}}(\omega )=\pi \left[ (1+\gamma _{q\mathbf{k}})\delta
\left( \omega -\omega _{q\mathbf{k}}\right) +(1-\gamma _{q\mathbf{k}})\delta
\left( \omega +\omega _{q\mathbf{k}}\right) \right].  \label{eq:PolarAns5}
\end{equation}%
Then, with some algebra, Eqs. (\ref{eq:MEboson1})-(\ref{eq:MEboson3}) yield,
for the unknown functional parameters $\gamma _{q\mathbf{k}}$ and $\omega _{q%
\mathbf{k}}$, the self-consistent equations%
\begin{eqnarray}
\gamma _{q\mathbf{k}}\omega _{q\mathbf{k}} &=&\varepsilon _{\mathbf{k}%
}+\varphi (0)n+\frac{1}{V}\sum_{\mathbf{k}^{\prime }}\varphi (\left\vert
\mathbf{k}-\mathbf{k}^{\prime }\right\vert )N_{q\mathbf{k}^{\prime }}, \\
\omega _{q\mathbf{k}}^{2} &=&L_{q}+\frac{2}{V}\sum_{\mathbf{k}^{\prime
}}\varphi (\left\vert \mathbf{k}-\mathbf{k}^{\prime }\right\vert )\lambda _{q%
\mathbf{k}^{\prime }}+  
\varepsilon _{\mathbf{k}}(\varepsilon _{\mathbf{k}}+2n\varphi (0))+\frac{2%
}{V}\sum_{\mathbf{k}^{\prime }}\varphi (\left\vert \mathbf{k}-\mathbf{k}%
^{\prime }\right\vert )N_{q\mathbf{k}^{\prime }}.   \notag \\
\label{eq:paramset}
\end{eqnarray}%
In these equations $N_{q\mathbf{k}}$ is formally given by Eq. (\ref%
{eq:Nk_bose}) and%
\begin{multline}
\lambda _{q\mathbf{k}}=\frac{1}{4V}\sum_{\mathbf{k}^{\prime }}\varphi
(\left\vert \mathbf{k}-\mathbf{k}^{\prime }\right\vert ) 
 \left\{ \frac{%
1+\gamma _{q\mathbf{k^{\prime }}}}{\widetilde{e}_{q}^{\hspace{3pt}\beta
\omega _{q\mathbf{k^{\prime }}}}-1}+\frac{1-\gamma _{q\mathbf{k^{\prime }}}}{%
\widetilde{e}_{q}^{\hspace{3pt}-\beta \omega _{q\mathbf{k^{\prime }}}}-1}%
\right\} + \\
+\frac{1}{2}\left\{ \frac{\left( \omega _{q\mathbf{k}}-\varepsilon _{\mathbf{%
k}}\right) \left( 1+\gamma _{q\mathbf{k}}\right) }{\widetilde{e}_{q}^{%
\hspace{3pt}\beta \omega _{q\mathbf{k}}}-1}-\frac{\left( \omega _{q\mathbf{k}%
}+\varepsilon _{\mathbf{k}}\right) \left( 1-\gamma _{q\mathbf{k}}\right) }{%
\widetilde{e}_{q}^{\hspace{3pt}-\beta \omega _{q\mathbf{k}}}-1}\right\} .
\end{multline}%
Of course, when $\gamma _{q\mathbf{k}}$ and $\omega _{q\mathbf{k}}$ are
known, for the $q$ GF related to $\Lambda _{q\mathbf{k}}\left( \omega
\right) $,%
\begin{equation}
G_{q\mathbf{k}}\left( \omega \right) =\int_{-\infty }^{+\infty }\frac{%
d\omega ^{\prime }}{2\pi }\frac{\Lambda _{q\mathbf{k}}\left( \omega ^{\prime
}\right) }{\omega -\omega ^{\prime }},
\label{Eq:134}
\end{equation}%
we will have%
\begin{equation}
G_{q\mathbf{k}}\left( \omega \right) =\frac{1}{2}\left\{ \frac{1+\gamma _{q%
\mathbf{k}}}{\omega -\omega _{q\mathbf{k}}}+\frac{1-\gamma _{q\mathbf{k}}}{%
\omega +\omega _{q\mathbf{k}}}\right\} 
=\frac{\omega +\varepsilon _{\mathbf{k}}+n\varphi (0)+\frac{1}{V}\sum_{%
\mathbf{k}^{\prime }}\varphi (\left\vert \mathbf{k}-\mathbf{k}^{\prime
}\right\vert )N_{q\mathbf{k}^{\prime }}}{\omega ^{2}-\omega _{q\mathbf{k}%
}^{2}}.  \label{eq:GqBoson}
\end{equation}%
As expected, Eq. (\ref{eq:GqBoson}) implies two poles $\omega =\pm \omega _{q%
\mathbf{k}}$ (with $\omega _{q\mathbf{k}}\geq 0$ ) on the $\omega $ real
axis and hence the parameter $\omega _{q\mathbf{k}}$ determines the $q$%
 energy spectrum of the undamped elementary excitations in the system,
formally given by
\begin{equation}
\omega _{q\mathbf{k}}^{2}=\left\{ L_{q}+\frac{2}{V}\sum_{\mathbf{k}^{\prime
}}\varphi (\left\vert \mathbf{k}-\mathbf{k}^{\prime }\right\vert )\lambda _{q%
\mathbf{k}^{\prime }}+\varepsilon _{\mathbf{k}}(\varepsilon _{\mathbf{k}%
}+2n\varphi (0)\right. 
+\left. \frac{2}{V}\sum_{\mathbf{k}^{\prime }}\varphi (\left\vert \mathbf{k}-%
\mathbf{k}^{\prime }\right\vert )N_{q\mathbf{k}^{\prime }})\right\} ^{\frac{1%
}{2}}.
\label{Eq:136}
\end{equation}%
In the polar ansatz (\ref{eq:PolarAns5}), to determine $\Lambda _{q\mathbf{k}%
}(\omega )$ and hence all the relevant $q$-thermodynamic quantities, the
problem remains to express the unknown higher order quantity $L_{q}$ in
terms of the parameters $\gamma _{q\mathbf{k}}$ and $\omega _{q\mathbf{k}}$.

This difficulty can be easily overcome if we limit ourselves to
explore the condensate phase (which is of main interest for our Bose model
in the high-density and low-$T$ limits) when a macroscopic population of the
zero-moment single-particle state takes place. Indeed, in this situation one
can obtain a formally exact expression for $L_{q}$ setting $\omega _{q\mathbf{k=0}%
}=0$ below a certain critical temperature \cite{Abrikosov:65,BiLsuperfluid},
to find, from Eq. (\ref{Eq:136}),
\begin{equation}
L_{q}=-\mu ^{2}+\mu \left( 2\varphi \left( 0\right) n+\frac{2}{V}\sum_{%
\mathbf{k}^{\prime }}\varphi (k^{\prime })N_{q\mathbf{k}^{\prime }}\right)
- \frac{2}{V}\sum_{\mathbf{k}^{\prime }}\varphi (k^{\prime })\lambda _{q%
\mathbf{k}^{\prime }},
\end{equation}%
where $N_{q\mathbf{k}}$ and $\lambda _{q\mathbf{k}}$ are given by Eqs. (\ref%
{eq:nksegnato}) and (\ref{eq:lambdakmin}).

Then, in the condensate state the $q$ energy spectrum of the elementary
excitations can be written in the form%
\begin{equation}
\omega _{q\mathbf{k}}=\left\{ \frac{k^{4}}{4}+k^{2}\left[ \frac{1}{V}\sum_{%
\mathbf{k}^{\prime }}\varphi (\left\vert \mathbf{k}-\mathbf{k}^{\prime
}\right\vert )N_{q\mathbf{k}^{\prime }}+
 -\overset{\overset{}{}}{(\mu -n\varphi (0))}\right] +\overset{%
\overset{}{}}{\Omega _{q\mathbf{k}}}\right\} ^{\frac{1}{2}},
\label{eq:omegak1}
\end{equation}%
where%
\begin{equation}
\Omega _{q\mathbf{k}}=\frac{2}{V}\sum_{\mathbf{k}^{\prime }}\left[ \varphi
\left( \left\vert \mathbf{k}-\mathbf{k}^{\prime }\right\vert \right)
-\varphi \left( \left\vert \mathbf{k}^{\prime }\right\vert \right) \right]
\left( \lambda _{q\mathbf{k}^{\prime }}-\mu N_{q\mathbf{k}^{\prime }}\right)
.
\end{equation}%
Of course, Eq. (\ref{eq:omegak1}) does not provide the explicit $q$-dispersion
relation as a function of $T$ and $n$ since $N_{q\mathbf{k}}$ and
$\lambda_{q\mathbf{k}}$ contain $\omega _{q\mathbf{k}}$ itself and
$\gamma _{q\mathbf{k}}$ which have to be still determined as solutions of the closed
system of equations (132) and (\ref{eq:omegak1}).

In general, this problem is difficult to solve and, as for the extensive
counterpart \cite{kalashnikov:73,Caramico:80}, one is forced to consider asymptotic
regimes for obtaining explicit results or to use numerical calculations.
Obviously, solving numerically the previous closed self-consistent set of coupled $q$ MEs,
by variation of $T$ and $q$, constitutes a formidable tour de force (beyond the purposes of the present paper) which requires a separate study.
Here, to avoid obscuring complicated calculations, we follow the procedure
already used for the extensive case \cite{Caramico:80} to obtain,
in a natural way, predictions for the condensate state assuming
\begin{equation}
\frac{N-N_{q0}}{N}=\frac{n-n_{q0}}{n}\ll 1,  \label{eq:condA}
\end{equation}%
and%
\begin{equation}
\frac{\left\vert \Delta \mu \right\vert }{n\varphi (0)}\ll 1,
\label{eq:condB}
\end{equation}%
where $\Delta \mu =\mu -n\varphi (0)$.

These inequalities, already used in the previous subsection just as the
basic conditions for the validity of the Bogoliubov approximation, must be
consistently satisfied at the end of calculations.

It is easy to check that under conditions (\ref{eq:condA}) and (\ref{eq:condB}),
with $\mu \simeq n\varphi (0)$ and $n_{q0} \simeq n$
 to leading order, the $q$ MEs simplify to \cite{nota:exp1}
\begin{equation}
\left\{
\begin{array}{l}
\gamma _{q\mathbf{k}}\omega _{q\mathbf{k}}\approx k^{2}/2+n\varphi (k) \\
\omega _{q\mathbf{k}}^{2}\approx \left[ \omega _{\mathbf{k}}^{(B)}\right]
^{2}+\Omega _{q\mathbf{k}}^{(0)}.%
\end{array}%
\right.  \label{eq:MEparam}
\end{equation}%
Here%
\begin{equation}
\omega _{\mathbf{k}}^{(B)}=\left\{ \frac{k^{4}}{4}+nk^{2}\varphi (\mathbf{k}%
)\right\} ^{\frac{1}{2}},  \label{eq:omegaBk}
\end{equation}%
is the Bogoliubov spectrum, and
\begin{multline}
\Omega _{q\mathbf{k}}^{(0)}=\left. \Omega _{q\mathbf{k}}\right\vert _{\mu
\simeq n\varphi (0)} \simeq \\
\simeq \frac{n}{V}\sum_{\mathbf{k}^{\prime }}\varphi \left( k^{\prime
}\right) \left[ \varphi \left( \left\vert \mathbf{k}-\mathbf{k}^{\prime
}\right\vert \right) -\varphi \left( k^{\prime }\right) \right] 
 \left\{ 1+\left[ \frac{1}{\widetilde{e}_{q}^{\hspace{3pt}\beta \omega
_{\mathbf{k}^{\prime }}^{\left( B\right) }}-1}+\frac{1}{\widetilde{e}_{q}^{%
\hspace{3pt}-\beta \omega _{\mathbf{k}^{\prime }}^{\left( B\right) }}-1}%
\right] \right\} + \\
+\frac{1}{V}\sum_{\mathbf{k}^{\prime }}\left[ \varphi \left( \left\vert
\mathbf{k}-\mathbf{k}^{\prime }\right\vert \right) -\varphi \left( k^{\prime
}\right) \right] \left( \omega _{\mathbf{k}^{\prime }}^{(B)}-\gamma _{%
\mathbf{k^{\prime }}}\frac{k^{\prime }}{2}\right) 
 \left[ \frac{1}{\widetilde{e}_{q}^{\hspace{3pt}\beta \omega _{\mathbf{%
k}^{\prime }}^{\left( B\right) }}-1}-\frac{1}{\widetilde{e}_{q}^{\hspace{3pt}%
-\beta \omega _{\mathbf{k}^{\prime }}^{\left( B\right) }}-1}\right] ,
\end{multline}%
provides an estimate of the correction to the Bogoliubov spectrum
where the inequality $%
\left\vert \Omega _{q\mathbf{k}}^{(0)}\right\vert \ll \left[ \omega _{
\mathbf{k}}^{(B)}\right] ^{2}$ has to be checked at the end of calculations. Of
course, from Eq. (\ref{eq:MEparam}), we have for the parameter $\gamma _{q\mathbf{k}}$%
\begin{equation}
\gamma _{q\mathbf{k}}\simeq \frac{(k^{2}/2+n\varphi \left( k\right) )}{%
\omega _{\mathbf{k}}^{(B)}}\left\{ 1-\frac{\Omega _{q\mathbf{k}}^{(0)}}{2%
\left[ \omega _{\mathbf{k}}^{(B)}\right] ^{2}}\right\}.   \label{eq:gamma1k}
\end{equation}
Thus, the original $q$-moment problem is solved and all the $q$-thermodynamic
properties for the condensate region under the conditions (\ref{eq:condA}) and (\ref{eq:condB}),
and in particular the $q$ depletion of the condensate
$n-n_{q0}=\sum_{\mathbf{k}\neq0} N_{q\mathbf{k}}$, follow immediately from
the spectral relations (\ref{eq:nksegnato}), (\ref{Eq:129}), and (\ref{Eq:134}),
with $\Lambda_{q\mathbf{k}}(\omega)$ given by the two-$\delta$-function representation (\ref{eq:PolarAns5}). All
the relevant expressions in terms of $\omega_{q\mathbf{k}}$ and $\gamma_{q\mathbf{k}}$
at the Bogoliubov approximation level can be easily derived or also obtained from Ref. \cite{Caramico:80}
with $e^{x}$ replaced by $\widetilde{e}_{q}^{\hspace{3pt}x}$. These are quite
cumbersome and not particularly instructive and hence they will not be reported here for brevity reasons.

It is remarkable that, by a straightforward extension to the case $q\neq 1$ of the
procedure used in Ref. \cite{Caramico:80} for the extensive case, one can
systematically estimate, at least in principle, the corrections to the
predictions of the Bogoliubov approximation for any plausible potential
$\varphi \left( k\right) $.

For the high-density Bose model under study in the regime of interest,
the parameters $\omega_{q\mathbf{k}}$ and $\gamma_{q\mathbf{k}}$ in all the summands can
be replaced by their leading expressions
$\omega_{\mathbf{k}}^{(B)}$ and
$\gamma_{\mathbf{k}}^{(B)}=[k^{2}/2+n\varphi(k)]/\omega _{\mathbf{k}}^{(B)}$
and the basic equations become essentially identical to that obtained
in Sec. \ref{sec_5.2}. In particular, with $(n\gamma)^{1/2}\gg1$ and $\beta\gg1$,
all the results (\ref{eq:DEP1q})-(\ref{eq:Cvq}) near to the extensive regime
are easily reproduced and the conditions (\ref{eq:condA})
and (\ref{eq:condB}) are found to be consistently satisfied.

As a conclusion, we make some comments which may be of
practical interest for future calculations. For the high-density Bose model
defined in Sec. \ref{sec_5.1}, and possibly for other models, the $q$ SDM
offers some advantages with respect to the method used in Sec. \ref{sec_5.2} based
on the Bogoliubov scenario. Adopting the $q$ SDM one avoids the \emph{a priori} replacement $%
a_{0}^{\dagger }$, $a_{0}\rightarrow N_{q0}^{1/2}$ in the original
Hamiltonian and hence some consequent conceptual difficulties \cite%
{Ettouhami:07} in both the extensive and nonextensive cases. Besides, it
allows one to obtain, in a systematic way, the corrections to the Bogoliubov
predictions. Finally, the procedure may be adapted to other more realistic
situations provided that the conditions (\ref{eq:condA}) and (\ref{eq:condB})
are consistently satisfied at the end of calculations.


\section{Concluding remarks\label{sec_6}}

In this paper we have extended the general formalism of two-time GFs \cite%
{Tyablikov:67,Zubarev:74,kalashnikov:73} to nonextensive quantum statistical
mechanics in the OLM representation \cite{martinez:00}. Particular attention
has been devoted to the spectral properties and to the concept of SDs,
which is expected to play an important role in explicit
calculations also in nonextensive quantum many-body problems. Besides, we
have presented the $q$ EMM and the
nonextensive version of the less known SDM \cite{kalashnikov:73} for a
direct calculation of the $q$ GFs and $q$ SDs, respectively.
A remarkable feature is that
these methods should allow one, at least in principle, to explore on the same
footing the nonextensivity effects for a wide variety of quantum many-body
systems overcoming the \emph{a priori} knowledge of the $q$ partition function and
hence of the $q$ free energy. Unfortunately, in contrast to the extensive case
\cite{Tyablikov:67,Zubarev:74,kalashnikov:73}, the $q$ CFs of two generic
operators $A$ and $B$ cannot be expressed in a simple way, as the $q$ GFs,
in terms of the $q$ SD $\Lambda _{qAB}(\omega )$ defined in terms of the
same operators. This introduces in the formalism some intrinsic difficulties
which, in view of the present still limited experience, can be overcome only
under appropriate constraints such as, for instance, close to the extensive
regime. Under the restrictive condition (\ref{eq:condition}), the
calculation of $q$ CFs is sensibly simplified and we have shown how the
formalism works in exploring the $q$-induced
nonextensivity effects on the low-temperature properties of the
model (78)-(80) for the high-density Bose gas with strong attraction between the
particles \cite{Babichenko:73}.

In any case, it is desirable to test again the effectiveness of the $q$ EM
and $q$ SD methods for other many-body problems in nonextensive quantum
statistical mechanics. A preliminary study along this direction can be found
in Ref. \cite{Cavallo06} where we have applied the $q$ SDM to an isotropic
Heisenberg model with long-range exchange interactions.

We wish to stress again that, in the context of practical calculations, the
crucial problem remains to understand how one can calculate the relevant $q$%
 CFs and the related $q$-thermodynamic quantities by using the general
formulas (\ref{eq:abq}) and (\ref{eq:baq}) beyond the limiting condition (\ref%
{eq:condition}).

In conclusion, in light of the great experience acquired in the
extensive statistical mechanics \cite{Abrikosov:65,Tyablikov:67,
Zubarev:74,kadanoff:62,mahan:90,majlis:00,kalashnikov:73,bogolyubov:63,cavallo:05, Abrikosov:91}
, we believe that the $q$ GFs technique and, in particular, the $q$ SDM may
constitute a powerful tool of investigation also in nonextensive many-body
theory.


\appendix

\section{Two-time Green's functions and spectral density method in
nonextensive classical thermostatistics \label{app_A}}

As mentioned in the Introduction, the pioneering framework of the two-time
GF method in extensive classical statistical mechanics by Bogoliubov and
Sadovnikov \cite{bogolyubov:63} has opened the concrete possibility to
describe classical and quantum many-body systems on the same footing.
Besides, in many physical situations (when the quantum effects are
negligible), the use of the classical formalism may offer substantial
advantages especially from the computational point of view because in the
calculations one handles only functions and not operators.

Recently, the two-time GF technique and SDM have been formulated in
nonextensive classical thermostatistics, within the OLM framework, in two
our papers \cite{cavallo:01,cavallo:02} and conveniently applied to the
ferromagnetic Heisenberg chain. For completeness, in this appendix,
we review the basic equations to underline the main differences with the
corresponding quantum case which has been the subject of the present article.

First, we note that the basic ingredients of the quantum Tsallis
thermostatistics shortly reviewed in Sec. \ref{sec_2} remain formally valid
for the classical framework. Here, $\rho =\rho (\mathbf{q},\mathbf{p})$
denotes the probability distribution defined in the space phase $\Gamma $, $%
\mathbf{q}=\left\{ q_{1},...,q_{\mathcal{N}}\right\} $ and $\mathbf{p}%
=\left\{ p_{1},...,p_{\mathcal{N}}\right\} $ are the generalized coordinates,
and Tr(...) stands for $\int (...)d\Gamma =\int \prod_{i=1}^{\mathcal{N}%
}dq_{i}dp_{i}$, for a classical system with Hamiltonian $H(\mathbf{q},%
\mathbf{p})$ and $\mathcal{N}$ degrees of freedom. As for the quantum case, also
in this appendix we adopt the classical OLM representation [see Eqs. (\ref%
{eq:constraint1})-(\ref{eq:exOLM2})].

\subsection{Classical two-time $q$ Green's functions and $q$-spectral density
\label{app_A.1}}

In classical nonextensive thermostatistics, the two-time $q$ GFs involving
two classical observables are defined by \cite{cavallo:01}
\begin{equation}
G_{qAB}^{(\nu )}\left( t,t^{\prime }\right) = \left\langle \left\langle
A(t);B(t^{\prime })\right\rangle \right\rangle _{q}^{(\nu )}  
= \theta _{\nu }\left( t-t^{\prime }\right) \left\langle \left\{
A(t),B(t^{\prime })\right\} \right\rangle _{q},  \label{eq:GFclassic}
\end{equation}%
where the set of Poisson brackets $\left\{...,...\right\} $ replaces the quantum
commutator $-i\left[ ...,...\right] _{\eta }$ and the $q$ expectation value
is taken over the phase space within the OLM spirit. Here, $X(t)$ (with $%
X\equiv A,B$) represents a dynamical variable depending on time through the
canonical coordinates with $X(t)=X(\mathbf{q}(t),\mathbf{p}(t))=X(e^{iL_{H}t}%
\mathbf{q}(0),e^{iL_{H}t}\mathbf{p}(0))=e^{iL_{H}t}X(0)$, where $%
L_{H}=i\left\{ H,...\right\} $ is the \textit{Liouville }operator. Hence, $%
e^{iL_{H}t}$ acts as a classical time-evolution operator which transforms
the dynamical variable $X(0)\equiv X$ at initial time $t=0$ into the
variable $X(t)$ at arbitrary time $t$, satisfying the Liouville equation $%
dX(t)/dt=iL_{H}X(t)$.

The main difference of the quantum and classical $q$ GF methods lies in
the relation between the $q$ GFs and the $q$ CFs.
Actually, starting from the definitions of
Poisson's brackets and $q$ mean value, it can be shown that \cite{cavallo:01}
\begin{equation}
G_{qAB}^{(\nu )}(t,t^{\prime })=q\beta \theta _{\nu }(t-t^{\prime })\frac{d}{%
dt}F_{qA_{q}B_{q}}(t,t^{\prime }),  \label{eq:GqAB_FqAqBq}
\end{equation}%
connecting $G_{qAB}^{(\nu )}(t,t^{\prime })$ to the new generalized two-time
$q$-CF
\begin{equation}
F_{qA_{q}B_{q}}(t,t^{\prime })=\left\langle A_{q}(t)B_{q}(t^{\prime
})\right\rangle _{q},  \label{eq:FqAqBq}
\end{equation}%
of the two $q$-dynamical variables $A_{q}$ and $B_{q}$ defined (working in
the canonical ensemble) by:
\begin{equation}
X_{q}=\frac{X}{\sqrt{C_{q}}},\text{ \ \ }C_{q}=1-\beta (1-q)(H-U_{q})>0.
\end{equation}%
Unfortunately, it is not possible to derive a direct relation between $%
G_{qAB}^{(\nu )}(t,t^{\prime })$ and $F_{qAB}(t,t^{\prime })$ as happens
in the extensive classical case \cite{cavallo:05}. Nevertheless, using the
series representation

\begin{equation}
\frac{1}{1-x}=1+2\sum_{n=1}^{\infty }J_{n}(nx),
\end{equation}%
where $J_{n}(z)$ are the Bessel functions
\begin{equation}
J_{n}(z)=\sum_{m=0}^{\infty }\frac{(-1)^{m}\left( \frac{1}{2}z\right) ^{2m+n}%
}{m!\Gamma \left( n+m+1\right) },
\end{equation}%
with $x=(1-q)\beta (H-U_{q})$, we can formally write the expansion
\begin{equation}
G_{qAB}^{(\nu )}(t,t^{\prime }) =q\beta \theta _{\nu }(t-t^{\prime })\frac{d%
}{dt}\left[ \left\langle A(t)B(t^{\prime })\right\rangle _{q} 
 +\sum_{n,m=1}^{\infty }C_{q}^{n,m}(\beta )\left\langle
A(t)B(t^{\prime })(H-U_{q})^{2m+n}\right\rangle _{q}\right] ,
\end{equation}
where
\begin{equation}
C_{q}^{m,n}(\beta )=\frac{2(-1)^{m}[\frac{n}{2}\beta (1-q)]^{2m+n}}{m!\Gamma
(n+m+1)}.
\end{equation}%
With the restriction $|x|<1$ the simplest series $\left( 1-x\right)
^{-1}=\sum_{n=0}^{\infty }x^{n}$ can be used to obtain the formula
\begin{equation}
G_{qAB}^{(\nu )}(t,t^{\prime }) =q\beta \theta _{\nu }(t-t^{\prime })\frac{d%
}{dt}\left[ \overset{}{\left\langle A(t)B(t^{\prime })\right\rangle }%
_{q} 
 +\sum_{n=1}^{\infty }\beta ^{n}(1-q)^{n}\left\langle A(t)B(t^{\prime
})(H-U_{q})^{n}\right\rangle _{q}\right] .  \label{eq:GqE2}
\end{equation}
In particular, we have
\begin{equation}
G_{qAB}^{(\nu )}(t,t^{\prime })\approx q\beta \theta _{\nu }(t-t^{\prime })%
\frac{d}{dt}\langle A(t)B(t^{\prime })\rangle _{q},  \label{eq:GqE2approx}
\end{equation}%
under the condition $|x|=|\beta (1-q)(H-U_{q})|\ll 1$. This shows that only
in the lowest order may the $q$ GFs be directly related to $%
F_{qAB}(t,t^{\prime })$ as in the extensive case \cite{cavallo:05}. Any way, we
can improve systematically the approximation (\ref{eq:GqE2approx}), taking
into account the successive terms in the expansion (\ref{eq:GqE2}) relating $%
q$ GF to $q$ CFs of increasing order.

Assuming time-translational invariance, all the spectral properties derived
in Sec. \ref{sec_3.1} remain formally unchanged, but now the classical $q$ SD$
\Lambda _{qAB}\left( \omega \right) $ is defined by
\begin{equation}
\Lambda _{qAB}\left( \omega \right) =i\int_{-\infty }^{+\infty }d\tau
e^{i\omega \tau }\left\langle \left\{ A(\tau ),B\right\} \right\rangle _{q}.
\end{equation}%
Besides, from Eqs. (\ref{eq:GqAB_FqAqBq}) and (\ref{eq:FqAqBq}), one can show
that \cite{cavallo:01}
\begin{equation}
\left\langle A_{q}(\tau )B_{q}\right\rangle _{q}=\int_{-\infty }^{+\infty }%
\frac{d\omega }{2\pi }\frac{\Lambda _{qAB}\left( \omega \right)e^{-i\omega
\tau }}{q\beta \omega }.
\end{equation}%
Then, using the series representation (\ref{eq:GqE2}), we get
\begin{equation}
\int_{-\infty }^{+\infty }\frac{d\omega }{2\pi }\frac{\Lambda _{qAB}\left(
\omega \right) }{q\beta \omega }=\left\langle AB\right\rangle _{q} 
+\sum_{n,m=1}^{\infty }C_{q}^{n,m}(\beta )\left\langle
AB(H-U_{q})^{2m+n}\right\rangle _{q},  \label{eq:A20}
\end{equation}%
which corresponds to Eq. (\ref{eq:abq}) and allows us to connect the classical $%
q$ CFs with the related $q$ SD. In the limit $q\rightarrow 1$, Eq. (\ref%
{eq:A20}) reproduces consistently the well-known exact extensive result \cite%
{cavallo:05}

\begin{equation}
\lim_{q\rightarrow 1}\left\langle AB\right\rangle _{q}=\left\langle
AB\right\rangle =\int_{-\infty }^{+\infty }\frac{d\omega }{2\pi }\frac{%
\Lambda _{1AB}\left( \omega \right) }{\beta \omega }.
\end{equation}%
In order to derive the spectral decomposition for $\Lambda _{qAB}\left(
\omega \right) $, we now introduce a Hilbert space $\mathcal{S}_{q}$ of the
classical dynamical variables with a scalar product defined conveniently by
\cite{cavallo:05,deGennes:book}
\begin{equation}
\langle A|B\rangle _{q}=\widetilde{Z}_{q}\langle A^{\ast }B\rangle _{q},
\end{equation}%
with $\widetilde{Z}_{q}=\int d\Gamma \lbrack 1-\beta (1-q)(H(\mathbf{q},%
\mathbf{p})-U_{q})]^{\frac{q}{1-q}}$. In this space one can consider the
eigenvalue equation $L_{H}\Psi _{k}=\omega _{k}\Psi _{k}$ for the Hermitian
Liouville operator. It is also immediate to prove that, if $\Psi _{k}$ is an
eigenfunction of $L_{H}$ with (real) eigenvalues $\omega _{k}$, then $\Psi
_{k}^{\ast }$ is also an eigenfunction of $L_{H}$ with eigenvalue $-\omega
_{k}$. If we assume that $\{\Psi _{k}\}$ is a complete set of orthonormal
eigenfunctions, we can consider the expansions $A(\mathbf{q},\mathbf{p}%
)=\sum_{k}\langle \Psi _{k}^{\ast }|A\rangle _{q}\Psi _{k}^{\ast }(\mathbf{q}%
,\mathbf{p})$ and $B(\mathbf{q},\mathbf{p})=\sum_{k}\langle \Psi
_{k}|B\rangle _{q}\Psi _{k}(\mathbf{q},\mathbf{p})$. Bearing this in mind,
we can write \cite{cavallo:01}
\begin{equation}
\Lambda _{qAB}(\omega )=2\pi q\beta \omega \widetilde{Z}_{q}^{-1}\sum_{k}%
\langle \Psi _{k}|B_{q}\rangle _{q}\langle \Psi _{k}^{\ast }|A_{q}\rangle
_{q}\delta (\omega -\omega _{k}),  \label{eq:A23}
\end{equation}%
which is the desired classical $q$-spectral decomposition for $\Lambda
_{qAB}(\omega )$. In particular, if $B=A^{\ast }$, we have that $\Lambda
_{qAA^{\ast }}(\omega )$ is a real and positive-definite quantity. A
consequence of Eq. (\ref{eq:A23}) is that $G_{qAB}(\omega )$ and $\langle
A_{q}(t)B_{q}\rangle _{q}$ can be written as
\begin{equation}
G_{qAB}(\omega )=2\pi q\beta \widetilde{Z}_{q}^{-1}\sum_{k}\langle \Psi
_{k}|B_{q}\rangle _{q}\langle \Psi _{k}^{\ast }|A_{q}\rangle _{q}\frac{%
\omega _{k}}{\omega -\omega _{k}}  \label{eq:Gpol_cl}
\end{equation}%
and
\begin{equation}
\langle A_{q}(\tau )B_{q}\rangle _{q}=\widetilde{Z}_{q}^{-1}\sum_{k}\langle
\Psi _{k}|B_{q}\rangle _{q}\langle \Psi _{k}^{\ast }|A_{q}\rangle
_{q}e^{-i\omega _{k}\tau }.  \label{eq:Fpol_cl}
\end{equation}%
Thus, also in the classical case, the real poles of $G_{qAB}(\omega )$
represent the frequency spectrum of undamped oscillations.

\subsection{Methods of calculation for classical two-time $q$ Green's
functions and $q$-spectral density \label{app_A.2}}

\subsubsection{Classical equations-of-motion method \label{app_A.3.1}}

As in the quantum case, successive differentiations of Eq. (\ref%
{eq:GFclassic}) with respect to $\tau =t-t^{\prime }$, yield the infinite
hierarchies of classical coupled EMs in $\tau$ and $\omega$
representations:
\begin{equation}
\frac{d}{d\tau }\left\langle \left\langle \mathcal{L}_{H}^{m}A(\tau
);B\right\rangle \right\rangle _{q}^{(\nu )} = \delta \left( \tau \right)
\left\langle \left\{ \mathcal{L}_{H}^{m}A,B\right\} \right\rangle _{q} 
+\left\langle \left\langle \mathcal{L}_{H}^{m+1}A(\tau );B\right\rangle
\right\rangle _{q}^{(\nu )}\text{ \ \ }(m=0,1,2,...)
\label{eq:EMclassic_t}
\end{equation}
and
\begin{equation}
\omega \left\langle \left\langle \mathcal{L}_{H}^{m}A(\tau);B\right\rangle
\right\rangle _{q,\omega }^{(\nu )} =i\left\langle \left\{ \mathcal{L}%
_{H}^{m}A,B\right\} \right\rangle _{q} 
+i\left\langle \left\langle \mathcal{L}_{H}^{m+1}A(\tau);B\right\rangle
\right\rangle _{q,\omega }^{(\nu )}\text{ \ \ }(m=0,1,2,...),
\label{eq:EMclassic_w}
\end{equation}%
respectively, where $\mathcal{L}_{H}=iL_{H}$ and $\mathcal{L}_{H}^{m}A$
means $\mathcal{L}_{H}^{0}A=A$, $\mathcal{L}_{H}^{1}A=\left\{ A,H\right\} $,
$\mathcal{L}_{H}^{2}A=\left\{ \left\{ A,H\right\} ,H\right\} $ and so on.

At this stage, the considerations made in Sec. \ref{sec_4.2} for solving the
EMs in the quantum case apply for the classical one, too.

\subsubsection{Classical $q$-spectral density method \label{app_A.3.2}}

From the definition $\Lambda _{qAB}\left( \tau \right) =i\left\langle
\left\{ A(\tau ),B\right\} \right\rangle _{q}$ of the classical $q$ SD in
the $\tau $ space, successive derivatives with respect to $\tau $ yield%
\begin{equation}
\frac{d^{m}}{d\tau^{m} }\Lambda _{qAB}\left( \tau \right) =i\left\langle
\left\{ \mathcal{L}_{H}^{m}A(\tau),B\right\} \right\rangle _{q}\text{ \ }%
(m=0,1,2,...).
\end{equation}%
Then, proceeding as in the quantum case, we easily find%
\begin{equation}
\int_{-\infty }^{+\infty }\frac{d\omega }{2\pi }\omega ^{m}\Lambda
_{qAB}\left( \omega \right) =-(i)^{m-1}\left\langle \left\{ \mathcal{L}%
_{H}^{m}A,B\right\} \right\rangle _{q}\text{ \ } 
(m=0,1,2,...).  \label{eq:CSDM}
\end{equation}%
The quantity on the left-hand side will be called the $m$ moment of $\Lambda
_{qAB}\left( \omega \right)$, and relation (\ref{eq:CSDM}) can be seen
as an infinite set of exact MEs or sum rules for the classical $q$ SD.

As for the quantum counterpart, due to the possibility to evaluate the Poisson brackets and hence the $q$ averages on the right-hand side of Eq. (\ref{eq:CSDM}), the $m$ moments can be explicitly
calculated without \emph{a priori} knowledge of $\Lambda _{qAB}\left( \omega
\right) $. So one can consider the sequence of equations, Eq. (\ref{eq:CSDM}), as a
typical moment problem to determine the unknown function $\Lambda
_{qAB}\left( \omega \right) $ and hence all the related macroscopic $q$%
 quantities. At this stage, in view of the classical exact $q$-spectral
decomposition (\ref{eq:A23}) for $\Lambda _{qAB}\left( \omega \right) $, the
basic idea of the SDM \cite{kalashnikov:73} for classical $q$%
 thermostatistics does not differ from the quantum case and all the
considerations made in Sec. \ref{sec_4} about the possible functional
representations for the $q$ SD preserve their validity in the context of the
classical $q$ many-body theory.

\section{Calculation of $\mathbf{k}$ sums as $V\rightarrow \infty $ in the
high-density and low-temperature limits\label{app_B}}

For the ($T=0$)-sums over $\mathbf{k}$ in Eqs. (\ref{eq:DEPq})-(\ref{eq:ENERq}),
with $V^{-1}\sum_{\mathbf{k}\neq 0}(...)=(2\pi ^{2})^{-1}\int_{0}^{\infty
}dkk^{2}(...)$ as $V\rightarrow \infty $ (all the summands depend only on $%
k=\left\vert \mathbf{k}\right\vert $) and the transformation
$x=k(n\gamma )^{-1/4}$, we can write%
\begin{equation}
\frac{1}{2V}\sum_{\mathbf{k}\neq 0}\frac{f_{\mathbf{k}}-\omega _{\mathbf{k}}%
}{\omega _{\mathbf{k}}}=\frac{(n\gamma )^{\frac{3}{4}}}{\left( 2\pi \right)
^{2}}\int_{0}^{\infty }dx\frac{\left[ \frac{x^{2}}{2}-f(x)\right] ^{2}}{f(x)}%
,
\end{equation}%
\begin{equation}
\frac{1}{2V}\sum_{\mathbf{k}\neq 0}\left( f_{\mathbf{k}}-\omega _{\mathbf{k}%
}\right) =\frac{(n\gamma )^{\frac{5}{4}}}{\left( 2\pi \right) ^{2}}%
\int_{0}^{\infty }dx\left[ \frac{x^{2}}{2}-f(x)\right] ^{2},
\end{equation}%
\begin{equation}
\frac{1}{2V}\sum_{\mathbf{k}\neq 0}k^{2}\frac{f_{\mathbf{k}}-\omega _{%
\mathbf{k}}}{\omega _{\mathbf{k}}}=\frac{(n\gamma )^{\frac{5}{4}}}{\left(
2\pi \right) ^{2}}\int_{0}^{\infty }dx\frac{x^{2}\left[ \frac{x^{2}}{2}-f(x)%
\right] ^{2}}{f(x)},
\end{equation}%
where%
\begin{equation}
f(x) =\left\{ \frac{x^{4}}{4}+(n\gamma )^{\frac{1}{2}}x^{2}\left[ \frac{%
(n\gamma )^{-\frac{1}{2}}}{(n\gamma )^{-\frac{1}{2}}+x^{2}} 
\label{eq:B4} 
 -\frac{e^{-(n\gamma )^{1/2}x^{2}\frac{R^{2}}{4}}}{\eta +1}%
\right] \right\} ^{\frac{1}{2}}.
\end{equation}
Following Babichenko \cite{Babichenko:73}, with $(n\gamma )^{1/2}\gg 1$ one
can now neglect the term $(n\gamma )^{-1/2}$ with respect to $x^{2}$ and the
Gaussian part in Eq. (\ref{eq:B4}). This is quite legitimate since the basic
contribution to the previous integrals is made by $x\sim 1$.

Then, one finds%
\begin{equation}
\frac{1}{2V}\sum_{\mathbf{k}\neq 0}\frac{f_{\mathbf{k}}-\omega _{\mathbf{k}}%
}{\omega _{\mathbf{k}}}\simeq a(n\gamma )^{\frac{3}{4}},
\end{equation}%
\begin{equation}
\frac{1}{2V}\sum_{\mathbf{k}\neq 0}\left( f_{\mathbf{k}}-\omega _{\mathbf{k}%
}\right) \simeq \frac{4}{5}b(n\gamma )^{\frac{5}{4}},
\end{equation}%
\begin{equation}
\frac{1}{2V}\sum_{\mathbf{k}\neq 0}k^{2}\frac{f_{\mathbf{k}}-\omega _{%
\mathbf{k}}}{\omega _{\mathbf{k}}}\simeq \frac{2}{5}b(n\gamma )^{\frac{5}{4}%
},
\end{equation}%
where%
\begin{equation}
a=\frac{1}{\left( 2\pi \right) ^{2}}\int_{0}^{\infty }dx\frac{\left[ (1+%
\frac{x^{4}}{4})^{\frac{1}{2}}-\frac{x^{2}}{2}\right] ^{2}}{(1+\frac{x^{4}}{4%
})^{\frac{1}{2}}}=\frac{\sqrt{2}}{48\pi ^{5/2}}\Gamma ^{2}(\frac{1}{4}),
\end{equation}%
\begin{equation}
b=\frac{1}{\left( 2\pi \right) ^{2}}\int_{0}^{\infty }dx\frac{(1+\frac{x^{4}%
}{4})^{\frac{1}{2}}-\frac{x^{2}}{2}}{(1+\frac{x^{4}}{4})^{\frac{1}{2}}}=%
\frac{\sqrt{2}}{4\pi ^{5/2}}\Gamma ^{2}(\frac{3}{4}).
\end{equation}%
Concerning the $T$-dependent sums over $\mathbf{k}$ in Eqs. (\ref{eq:DEPq}%
)-(\ref{eq:ENERq}), in the low-temperature limit only small values of $k$
contribute to the integrals which contain exponential functions. So, with
the additional condition $(n\gamma )^{1/2}\gg 1$, all the integrals can be
exactly computed and we get, to leading order in $T$,%
\begin{equation}
\frac{1}{V}\sum_{\mathbf{k}\neq 0}\frac{\omega _{\mathbf{k}}}{e^{\beta
\omega _{\mathbf{k}}}-1}\simeq \frac{\pi ^{2}}{30}\left( \frac{\eta +1}{\eta
}\right) ^{\frac{3}{2}}\left( n\gamma \right) ^{-\frac{3}{2}}T^{4},
\end{equation}%
\begin{equation}
\frac{1}{V}\sum_{\mathbf{k}\neq 0}\frac{f_{\mathbf{k}}}{\omega _{\mathbf{k}}}%
\frac{1}{e^{\beta \omega _{\mathbf{k}}}-1}\simeq \frac{1}{12}\left( \frac{%
\eta +1}{\eta }\right) ^{\frac{1}{2}}\left( n\gamma \right) ^{-\frac{1}{2}%
}T^{2},
\end{equation}%
\begin{equation}
\frac{1}{V}\sum_{\mathbf{k}\neq 0}\frac{\omega _{\mathbf{k}}^{2}e^{\beta
\omega _{\mathbf{k}}}}{\left( e^{\beta \omega _{\mathbf{k}}}-1\right) ^{2}}%
\simeq \frac{2\pi ^{2}}{15}\left( \frac{\eta +1}{\eta }\right) ^{\frac{3}{2}%
}\left( n\gamma \right) ^{-\frac{3}{2}}T^{5},
\end{equation}%
\begin{equation}
\frac{1}{V}\sum_{\mathbf{k}\neq 0}\frac{f_{\mathbf{k}}e^{\beta \omega _{%
\mathbf{k}}}}{\left( e^{\beta \omega _{\mathbf{k}}}-1\right) ^{2}}\simeq
\frac{1}{6}\left( \frac{\eta +1}{\eta }\right) ^{\frac{1}{2}}\left( n\gamma
\right) ^{-\frac{1}{2}}T^{3},
\end{equation}%
\begin{equation}
\frac{1}{V}\sum_{\mathbf{k}\neq 0}\frac{\omega _{\mathbf{k}}^{2}\varphi
\left( k\right) e^{\beta \omega _{\mathbf{k}}}}{\left( e^{\beta \omega _{%
\mathbf{k}}}-1\right) ^{2}}\simeq \frac{2\pi ^{2}}{15}\left( \frac{\eta +1}{%
\eta }\right) ^{\frac{1}{2}}\gamma \left( n\gamma \right) ^{-\frac{3}{2}%
}T^{5},
\end{equation}%
\begin{equation}
\frac{1}{V}\sum_{\mathbf{k}\neq 0}\frac{k^{2}\varphi \left( k\right)
e^{\beta \omega _{\mathbf{k}}}}{\left( e^{\beta \omega _{\mathbf{k}%
}}-1\right) ^{2}}\simeq \frac{2\pi ^{2}}{15}\left( \frac{\eta +1}{\eta }%
\right) ^{\frac{3}{2}}\gamma \left( n\gamma \right) ^{-\frac{5}{2}}T^{5},
\end{equation}%
\begin{equation}
\frac{1}{V}\sum_{\mathbf{k}\neq 0}\frac{f_{\mathbf{k}}\omega _{\mathbf{k}%
}^{2}e^{\beta \omega _{\mathbf{k}}}}{\left( e^{\beta \omega _{\mathbf{k}%
}}-1\right) ^{2}}\simeq \frac{2\pi ^{2}}{15}\left( \frac{\eta +1}{\eta }%
\right) ^{\frac{1}{2}}\left( n\gamma \right) ^{-\frac{1}{2}}T^{5}.
\end{equation}%
Inserting the previous results into Eqs. (\ref{eq:DEPq})-(\ref{eq:ENERq}), one
immediately obtains expressions (\ref{eq:DEP1q})-(\ref{eq:ENER1q})
for the depletion of the condensate, the chemical potential, the $q$ internal
energy density, and hence the $q$ specific heat (\ref{eq:Cvq}) as a
function of $T$ and $n$ in the high-density and low-temperature limits.



\begin{thebibliography}{99}
\bibitem{nota:app.A} To be compared with the classical framework presented
in Appendix A.
\bibitem{nota:commutatori} In the following we will put $[..,..]_{-1}\equiv
\lbrack ..,..]_{-}$ and $[..,..]_{+1}\equiv \lbrack ..,..]_{+}$ for formal
semplicity.
\bibitem{nota:exp1} We omit here the inessential mathematical details which do not depend on the
underlying presence of the parameter $q$ and follow trivially from those presented
in Ref. \cite{Caramico:80} with the usual exponential function $e^{x}$
replaced by $\widetilde{e}_{q}^{\hspace{3pt}x}$.
\end{thebibliography}
\end{document}